\documentclass[fleqn,usenatbib]{mnras}

\usepackage{newtxtext,newtxmath}
\usepackage[T1]{fontenc}
\usepackage{ae,aecompl}
\usepackage{graphicx}	
\usepackage{amsmath}	
\usepackage{svg}
\usepackage{gensymb}
\usepackage{xspace}
\usepackage[export]{adjustbox} 
\usepackage{xcolor}
\usepackage{orcidlink}
\usepackage{mathtools}
\DeclarePairedDelimiter\abs{\lvert}{\rvert}%

\newcommand{\phn}{$\hphantom{0}$}
\newcommand{\phs}{$\hphantom{-}$}
\newcommand{\Gaia}{{\sl Gaia}\xspace}
\newcommand{\Kepler}{{\sl Kepler}\xspace}
\newcommand{\perpix}{\mbox{pixel$^{-1}$}}
\newcommand{\kms}{\mbox{km\,s$^{-1}$}}
\newcommand{\masyr}{\hbox{mas\,yr$^{-1}$}}
\newcommand{\Kp}{\mbox{$K^{\prime}$}}
\newcommand{\Kc}{\mbox{$K_{\rm cont}$}}
\newcommand{\Mtot}{\mbox{$M_{\rm tot}$}}
\newcommand{\Lbol}{\mbox{$L_{\rm bol}$}}
\newcommand{\Teff}{\mbox{$T_{\rm eff}$}}

\newcommand{\istar}{\mbox{$i_{\star}$}}
\newcommand{\ipl}{\mbox{$i_{\rm pl}$}}
\newcommand{\phimax}{\mbox{$\phi_{\rm max}$}}

\title[Kepler Planet--Binary Mutual Inclinations]{Orbital Architectures of Planet-Hosting Binaries II. Low Mutual Inclinations Between Planetary and Stellar Orbits}

\author[T. J. Dupuy et al.]{Trent J.~Dupuy$^{1}$\thanks{E-mail: tdupuy@roe.ac.uk}\orcidlink{0000-0001-9823-1445},
Adam L.\ Kraus$^{2}$\orcidlink{0000-0001-9811-568X},
Kaitlin M.\ Kratter$^{3}$\orcidlink{0000-0001-5253-1338},
Aaron C.\ Rizzuto$^{2}$\thanks{51 Pegasi b Fellow}\orcidlink{0000-0001-9982-1332},
Andrew W.\ Mann$^{4}$\orcidlink{0000-0003-3654-1602}, \newauthor
Daniel Huber$^{5}$\orcidlink{0000-0001-8832-4488},
and Michael J.\ Ireland$^{6}$\orcidlink{0000-0002-6194-043X}
\\
$^{1}$Institute for Astronomy, University of Edinburgh, Royal Observatory, Blackford Hill, Edinburgh, EH9 3HJ, UK\\
$^{2}$Department of Astronomy, The University of Texas at Austin, 2515 Speedway C1400, Austin, TX 78712, USA\\
$^{3}$Department of Astronomy, University of Arizona, 933 N.\ Cherry Ave, Tucson, AZ 85721, USA\\
$^{4}$Department of Physics and Astronomy, University of North Carolina at Chapel Hill, Chapel Hill, NC 27599-3255, USA\\
$^{5}$Institute for Astronomy, University of Hawaii, 2680 Woodlawn Drive, Honolulu, HI 96822, USA\\
$^{6}$Research School of Astronomy \& Astrophysics, Australian National University, Canberra ACT 2611, Australia
}

\pubyear{2022}

\begin{document}
\label{firstpage}
\pagerange{\pageref{firstpage}--\pageref{lastpage}}
\maketitle

\begin{abstract}
Planet formation is often considered in the context of one circumstellar disk around one star. Yet stellar binary systems are ubiquitous, and thus a substantial fraction of all potential planets must form and evolve in more complex, dynamical environments. We present the results of a five-year astrometric monitoring campaign studying 45 binary star systems that host Kepler planet candidates. The planet-forming environments in these systems would have literally been shaped by the binary orbits that persist to the present day. Crucially, the mutual inclinations of star-planet orbits can only be addressed by a statistical sample. We describe in detail our sample selection and Keck/NIRC2 laser guide star adaptive optics observations collected from 2012 to 2017. We measure orbital arcs, with a typical accuracy of $\sim$0.1\,\masyr, that test whether the binary orbits tend to be aligned with the edge-on transiting planet orbits. We rule out randomly-distributed binary orbits at 4.7$\sigma$, and we show that low mutual inclinations are required to explain the observed orbital arcs. If the stellar orbits have a field binary-like eccentricity distribution, then the best match to our observed orbital arcs is a distribution of mutual inclinations ranging from 0\degree--30\degree. We discuss the implications of such widespread planet-binary alignment in the theoretical context of planet formation and circumstellar disk evolution.
\end{abstract}


\section{Introduction}

The majority of all Sun-like stars form with at least one stellar companion \citep[e.g.,][]{1969JRASC..63..275H,2008ApJ...679..762K,2010ApJS..190....1R,2017ApJS..230...15M}. Some of the defining physical characteristics of multiple-star systems are encoded in their orbital elements, with eccentricity ($e$) and semimajor axis ($a$) tracing the angular momentum and energy of a binary. The underlying distributions of such elements are well established (see \citealp{2013ARA&A..51..269D} for a review), with $a$ following a log-normal distribution \citep[e.g.,][]{1991A&A...248..485D} and the $e$ distribution being approximately flat for all but the shortest-period binaries \citep[e.g.,][]{2006ApJ...651.1151A}. 

From a theoretical perspective, binaries are expected to have a profound dynamical influence over the many stages of planetary formation and evolution. For example, binary companions truncate circumstellar disks \citep[e.g.,][]{1994ApJ...421..651A,2010ApJ...710..462A,2015ApJ...799..147J} and also dynamically stir planetesimals \citep[e.g.,][]{2007ApJ...660..807Q,2007ApJ...666..436H,2010MNRAS.403L..64P,2015ApJ...798...69R}. Furthermore, even if planets can be formed, secular evolution of the orbits and even stellar evolution can drive systems through unstable states that destroy them on $\sim$Myr to $\sim$Gyr timescales \citep[e.g.,][]{1999AJ....117..621H,2012ApJ...753...91K,2013Natur.493..381K}. 
In spite of these dynamical barriers, ground-based exoplanet surveys were the first to show that some binary systems do host planets. A handful of giant planets have been identified in nearby short-period binary systems \citep{2000A&A...354...99Q,2003ApJ...599.1383H,2008A&A...479..271C,2015ApJ...815...32K}, though it has been suggested that they might result from small-$N$ dynamical interactions rather than in situ formation \citep[e.g.,][]{2006ApJ...652.1694P}. In addition, circumbinary gas giants have been found in the \Kepler\ survey \citep{2011Sci...333.1602D}, perhaps as frequently as for single stars \citep[e.g.,][]{2014MNRAS.444.1873A,2016ApJ...831...96L, 2019A&A...624A..68M}. Wide binary companions have been found to many planet hosts \citep[e.g.,][]{2005A&A...440.1051M,2007A&A...474..273E,2019MNRAS.490.5088M}, implying that they too are common exoplanet hosts \citep{2019arXiv191201699M}. 

Crucially, however, binary pairs of $10 \lesssim a \lesssim 100$\,AU are observed to be hostile sites for planets to grow \citep[e.g.,][]{2014ApJ...783....4W,2016AJ....152....8K,2019arXiv191201699M,2020AJ....159...19Z}. Suppression of planet occurrence in these systems is consistent with theoretical modelling of the binaries' origins. While both turbulent core fragmentation and disk fragmentation can produce the widest binaries in this set, disk fragmentation is preferred for smaller separations \citep{2019ApJ...887..232L,2020MNRAS.491.5158T}. Disk fragmentation is expected to be disruptive to planet formation for a host of reasons, including the siphoning off of mass by the newly formed secondary star and the truncation of the individual protoplanetary disks. Indeed, roughly two thirds of close binary pairs ($\lesssim$50\,AU) appear to dispel their disks promptly \citep{2009ApJ...696L..84C,2012ApJ...745...19K,2019ApJ...878...45B}, though various combinations of circumprimary, circumsecondary, and circumbinary disks have been found to persist \citep[e.g.,][]{Prato:1997qv,2016ApJ...817..164R,2017ApJ...845..161A,2021ApJ...912....6C}.

Testing if particular orbital parameters are related to promotion or suppression of planet formation, and why, is a crucial step in understanding these ubiquitous, potentially planet hosting, environments. One prediction of disk fragmentation models is the alignment of the binary orbital plane with that of both protoplanetary disks. Thus when planets do form in such systems, they would also generally be expected to be aligned at least at birth, even though subsequent evolution might excite inclinations \citep{JenningsChiang:2021}. \citet{2017ApJ...835L..28M} also show that polar (orthogonal) orientations are dynamically stable for circumbinary protoplanetary disks, and indeed these have been observed over separations of $\sim$1--100\,AU \citep{2019ApJ...883...22C}. Observations have found protoplanetary disks within young binary systems with wide-ranging separations to be both aligned and misaligned with respect to each other \citep{1998ApJ...502L..65S,2004ApJ...600..789J} and with respect to the orbits of their binary systems \citep{2004ApJ...603L..45W,2013A&A...554A.110P,2014AJ....147..157S,2019AJ....157...71P}. This diversity is consistent with the dual formation channels of binaries; as the separation decreases from hundreds to tens of AU and less, the propensity for alignment is expected to increase. Reliable observational confirmation of this trend is still required \citep{2020A&A...642A.212J}.

The two observed trends of planet suppression in closer binaries and binary--disk alignment increasing toward smaller $a$ suggests that the orbital planes of stars and planets in systems that form planets would tend to be misaligned. Yet it is also possible for young misaligned disks to give rise to well-aligned planet-binary systems. Disk warping and damping can re-align the disk and binary orbital planes on timescales shorter than the disk dissipation, and thus planet formation, timescale \citep{Ogilvie:2000}. Acting in the opposite direction at later times, secular dynamical interactions can also misalign nearly coplanar systems after the disk disperses \citep{2017ApJ...835L..28M,Petrovich2020}. 

Observations of planet--binary mutual inclinations are rare outside of the circumbinary planets orbiting very short-period binaries that are broadly aligned \citep[e.g.,][]{2015ApJ...809...26W}. In the case of Kepler-444, a star hosting five transiting Mars-sized planets and a wide ($\approx$40\,AU) outer double-M-dwarf companion system, \citet{2016ApJ...817...80D} found that the companion orbit was consistent with being aligned with the planetary orbits. At a much wider scale, Kepler-25 and KOI-1803 both host planets and are separated by 2000\,AU but have an orbital inclination that is consistent with only a small $\approx$10$\degree$ misalignment \citep{2020ApJ...894..115P}. Yet such individual case studies cannot rule out arbitrarily misaligned orbits. A perfectly edge-on planetary system and edge-on binary could still be misaligned in the position angles of their orbital nodes (i.e., in the sky plane). 

The continuously extending maturity of the \Kepler\ planet sample now offers the opportunity for a population-level comparison of its (edge-on) planetary population to the motions of previously-discovered binary companions. In this paper, we present results from the first five years of our Keck adaptive optics (AO) astrometric monitoring of a sample of binary systems hosting planet candidates from \Kepler\ (KOIs). From these data we have measured precise orbital arcs and compared the observed motion with a wide range of underlying orbital parameter distributions. We demonstrate in the following that the observed distribution is most consistent with planet and stellar orbital planes tending to have low mutual inclinations. We then discuss the implications of such observed present-day mutual inclinations for planetary formation and survival in the dynamically complicated environments of multiple star systems.

\section{Observations}

\subsection{Sample selection  \label{sec:sample}}

The input sample for our orbit monitoring were the binaries reported in \citet{2016AJ....152....8K}. That survey in turn used the NASA Exoplanet Archive\footnote{\url{https://exoplanetarchive.ipac.caltech.edu/}} list of known KOIs as of 2013~Jul~31 for its input sample and prioritised targets based on their estimated photometric distances, resulting in an approximately volume-limited sample of 382 KOIs. We prioritised our astrometric follow up of the 252 KOIs with one or more companions based on the speed of expected orbital motion of the closest companion. For a face-on, circular orbit the angular orbital speed is $2\pi\rho/P$, where $\rho$ is the angular separation and $P$ is the orbital period.  We used the best available distances ($d$) and component masses ($M_{\rm pri}$ and $M_{\rm sec}$) at the time, as estimated by \citet{2016AJ....152....8K}, to compute semimajor axes ($a = \rho d$) and thereby orbital periods $P = \sqrt{a^3/(M_{\rm pri} + M_{\rm sec})}$. Of course, eccentricity, inclination, and the time of periastron passage relative to the current epoch will cause the true orbital motion to vary, but without prior knowledge of these parameters our best estimate for the orbit speed will still be proportional to  $\rho/P$, which scales as $\rho^{-1/2} d^{-3/2} (M_{\rm pri} + M_{\rm sec})^{1/2}$. As a secondary selection factor, we considered the likelihood that the companion was a background object based on the field contamination estimates done by \citet[][see their Figure~4]{2016AJ....152....8K}.

Due to the strong dependence on distance, the fastest orbit speed estimate for the whole input sample was 21\,\masyr\ for the nearby (38\,pc) system KOI-3158 (Kepler-444). In order to include KOI-0214, which had a relatively small estimated orbital period of 240\,yr, we considered systems with orbit speeds down to 1.9\,\masyr.  From the 65~systems meeting this estimated orbit speed cut, we excluded five systems because the companion was sufficiently faint and widely separated to be a likely background star (KOI-0387, KOI-0663, KOI-1515, KOI-1843, KOI-2704), as well as three faint companions that required the use of the NIRC2 coronagraph (KOI-0069, KOI-0314, KOI-0961). None of these companions are detected in \Gaia~EDR3 \citep{2021_GaiaEDR3_Summary}. One more system was excluded from further monitoring after a single follow-up epoch confirmed the closest companion was a background object (KOI-1174). Finally, we excluded KOI-3908 because we now believe that the original masking detection reported in \citet{2016AJ....152....8K} was a systematic caused by an unusual quasi-stable tip-tilt oscillation in the AO system. Our monitoring sample thus consisted of 55 KOIs.

\begin{table*}
\caption[]{Relative astrometry measurements of KOIs from our Keck/NIRC2 adaptive optics imaging and aperture-masking interferometry.} \label{tbl:keck}
\begin{tabular}{lcccccl}
\hline
System & \multicolumn{2}{c}{Observation epoch} & Separation & Position angle & $\Delta{m}$ & Filter \\
 & (UT) & (MJD) & (mas) & ($\degree$) & (mag) & \\ \hline
KOI-0001AB & 2012~Jul~6  & 56114.59 & $1105.1\pm0.4    $ & $136.351\pm0.020  $ & $ 2.351\pm0.000  $ & $\Kp     $ \\
KOI-0001AB & 2015~Jun~23 & 57196.55 & $1110.3\pm0.5    $ & $136.369\pm0.022  $ & $ 2.386\pm0.007  $ & $\Kp     $ \\
KOI-0001AB & 2016~Jun~16 & 57555.59 & $1112.7\pm0.5    $ & $136.385\pm0.022  $ & $ 2.347\pm0.008  $ & $\Kp     $ \\\hline
KOI-0042AB & 2012~May~6  & 56053.64 & $1667.2\pm0.7    $ & $ 35.534\pm0.020  $ & $ 1.854\pm0.020  $ & $\Kp     $ \\
KOI-0042AB & 2014~Jul~31 & 56869.36 & $1666.3\pm0.7    $ & $  35.61\pm0.04   $ & $  1.86\pm0.03   $ & $\Kc     $ \\
KOI-0042AB & 2015~Jul~27 & 57230.43 & $1665.2\pm0.8    $ & $ 35.592\pm0.024  $ & $ 1.905\pm0.015  $ & $\Kc     $ \\
KOI-0042AB & 2016~Jun~16 & 57555.57 & $1664.4\pm0.8    $ & $ 35.607\pm0.024  $ & $ 1.838\pm0.020  $ & $\Kc     $ \\\hline
KOI-0214AB & 2014~Aug~13 & 56882.30 & $  70.9\pm1.6    $ & $  196.2\pm1.3    $ & $  3.71\pm0.10   $ & $\Kp+9  $H \\
KOI-0214AB & 2015~Jul~21 & 57224.54 & $  65.1\pm2.8    $ & $  194.6\pm2.6    $ & $  3.49\pm0.20   $ & $\Kp+9  $H \\\hline
\end{tabular}
\smallskip
\begin{list}{}{}
\item[\sc Note---]The filter column includes a ``+ 9H'' next to the filter name to denote when a measurement was derived from non-redundant aperture masking interferometry using the 9-hole mask of NIRC2. Astrometric uncertainties for imaging results are computed as quadrature sum of the rms among measurements at a given epoch and the NIRC2 pixel scale and orientation uncertainties. For masking results we use the errors derived by our pipeline from the ensemble of interferograms at a given epoch.  We neglect the uncertainty in the distortion solution as is appropriate for binaries with small angular separations (see Section~\ref{sec:keck}).  Each $\Delta{m}$ uncertainty is simply the rms among measurements at a given epoch. \emph{(The full table is available at the end of this preprint.)}
\end{list}
\end{table*}

There are a number of systems that we have obtained follow-up observations on that we do not include in the following analysis. Six of the nineteen systems with masking-detected companions (KOI-0289, KOI-1316, KOI-1397, KOI-1833, KOI-1902, KOI-2036) have not been robustly recovered in our follow-up observations to date. In some cases this is clearly due to worse seeing compared to the original detection epochs, but it is also possible that orbital motion has made separations tighter, inside our detection limits, for some of these systems. 

The other group of objects that we exclude from our analysis are higher order multiple systems unresolved by \Kepler\ (KOI-0005, KOI-2626, KOI-2813, KOI-3497).\footnote{There are widely separated sources in \Gaia~EDR3 that seem to share common proper motion and parallax with KOI-1961, KOI-2059, and KOI-2733. We retained these candidate higher-order multiple systems as our analysis focuses on the inner planet-hosting binary.} In these systems we cannot readily determine which of the orbital planes is the correct one to compare to the planet's orbit because the identity of the host star is uncertain. (In binary systems where the host star identity is uncertain, there is only one stellar orbital plane to consider, eliminating this problem.) The analysis of these dynamically complex systems will be reserved for a future study.  

Therefore, of the 55 KOIs we have monitored, we report orbit measurements for the 45 binaries that have multi-epoch data to date.  As we show in the following, all of these 45 binaries are physically bound and none has any additional companions.

\subsubsection{False positive re-analysis \label{sec:fp}}

Given that our sample comprises entirely multiple star systems, and not all planets around them have been validated, there is a strong possibility for some of these KOIs to harbour false positives (FPs). On the other hand, FP classifications could be erroneous when due entirely to centroid offsets, as such offsets are actually expected for astrophysical true positives in binary systems. Out of our entire sample, the only KOI with an FP classification by NExScI for reasons other than centroid offsets is KOI-1961.02. We still include KOI-1961 in our sample because KOI-1961.01 is not classified as a FP and is confirmed by our own re-analysis of FP classifications, described as follows.

We used the primary star's \Teff, \Lbol, and [Fe/H] from \citet{2020AJ....159..280B} and the measured $\Delta{K}$ to estimate stellar properties (including mean densities) for both binary components using \texttt{isoclassify} \citep{2017ApJ...844..102H}. This procedure is essentially the same as we used in \citet{2016AJ....152....8K} but with an improved stellar classification method and a newer grid of MIST isochrones \citep{Choi16_MIST} as described in \citet{2020AJ....159..280B}. We then calculated expected stellar densities for each component from the measured transit durations, assuming uniform impact parameters over $(0,1-R_{\rm pl}/R_{\star})$ (i.e., excluding grazing transits) and a Rayleigh distribution with a mean of 0.05 for eccentricities \citep{vaneylen15}. We then compared the two stellar densities (stellar isochrone and transit duration) for each component of the binary and calculated a probability of the planet orbiting the primary from Monte-Carlo simulations, following the same method used in \citet{2016MNRAS.457.2877G}. We found that all KOIs had transit durations consistent with orbiting either the primary or the secondary.

Nine stars in our sample have planets that are more likely to orbit the secondary than primary. All but one are only marginal, with the probability of orbiting the primary between $p_{\rm prim} = 0.43$--0.50. KOI-3444.02 is the only planet that appears to be much more likely to orbit the secondary ($p_{\rm prim} < 10^{-4}$), while KOI-3444.03 and KOI-3444.04 are likely to orbit the primary ($p = 0.997$ and 0.89, respectively). The fourth planet in the system KOI-3444.04 is marginal between the primary and secondary ($p_{\rm prim} = 0.50$). We note that the KOI-3444.02 transit is V-shaped, which means it could be a larger planet orbiting KOI-3444A rather than a small planet orbiting KOI-3444B. We find this to be unlikely because KOI-3444.02 shows a centroid offset roughly matching the secondary separation, and it would be somewhat unusual to have three sub-Earth-sized planets with a nearby warm Neptune (or larger) planet. Regardless, KOI-3444 is the only system with possible evidence for three (rather than two) orbital planes of two independently edge-on planetary orbits plus the stellar orbit.

In summary, all of our targets contain at least one planet candidate that does not have any evidence of being a false positive, and none (except possibly KOI-3444) has any evidence that one of the planets orbits the fainter star in the system.

\subsection{Keck/NIRC2 LGS AO imaging \& masking \label{sec:keck}}

All of our observations were obtained using the facility camera NIRC2\footnote{\url{https://www2.keck.hawaii.edu/inst/nirc2/}} at the Keck~II Telescope on Maunakea, Hawai`i. We used the narrow camera mode of NIRC2 that provides the finest pixel sampling (10\,mas\,\perpix). We used the broadband \Kp\ filter ($\lambda_{\rm eff} = 2.108$\,\micron, ${\rm FWHM} = 0.352$\,\micron) for all targets except for the brightest ones ($K\loa9$\,mag) for which we used the narrow-band \Kc\ filter ($\lambda_{\rm eff} = 2.287$\,\micron, ${\rm FWHM} = 0.032$\,\micron). We typically used the laser guide star adaptive optics (LGS AO) system \citep{2006PASP..118..297W,2006PASP..118..310V}, and for brighter targets ($R\loa9$\,mag) we used natural guide star (NGS) AO \citep{2000PASP..112..315W}. For the tightest binaries with separations $\loa0\farcs1$, we obtained interferograms on NIRC2 by using the 9-hole aperture mask that is installed in one of the filter wheels of NIRC2.

Our data were collected during 36 distinct nights spanning 2012~May~6~UT to 2017~Jul~7~UT. This includes 18~nights for which we have previously published data \citep{2016AJ....152....8K}. We performed a homogeneous analysis of our imaging and aperture masking interferometry observations for the analysis we present here. For all data sets we use the reduction pipeline products described by \citet{2016AJ....152....8K}, e.g., for linearity corrections, flats, and darks, but we perform a new astrometric analysis of these calibrated images.

To derive astrometry from our imaging data, we used the same methods as described in \citet{2019AJ....158..174D}, which is the latest iteration of our NIRC2 astrometry pipeline developed in and improving on our previous work \citep[e.g.,][]{2008ApJ...689..436L,2016ApJ...817...80D,2017ApJS..231...15D}. Briefly, we derived $(x,y)$ positions for each star in NIRC2 detector coordinates by fitting either an empirical template PSF computed from the image itself using StarFinder \citep{2000A&AS..147..335D} or, for tighter binaries with insufficiently separated PSFs for such an approach, an analytic multi-component Gaussian PSF. 

For our masking data, we used the same data analysis pipeline as was used by \citet[][see Section~3.4 of that work]{2016AJ....152....8K}, which is largely based on the analysis methods described in \citet{2008ApJ...679..762K}. Briefly, the pipeline computed closure phases and visibilities from the interferograms, used the rms of measurements across individual interferograms taken on the same night as initial error bars, and fitted a binary model to them. The uncertainties were subsequently increased by adding a calibration error in quadrature so that the resulting reduced $\chi^2$ of the fit was 1.0.

The NIRC2 coordinates we found through PSF-fitting and interferomteric analysis were then transformed into angular separations and position angles (PAs) using the astrometric calibration of \citet[][]{2010ApJ...725..331Y} for data obtained prior to the realignment of the AO system on 2015~Apr~13 and \citet{2016PASP..128i5004S} for later epochs. These calibrations correct for nonlinear distortion and provide the pixel scale and a correction to the orientation of NIRC2 provided in the image headers. At this stage we also accounted for differential aberration and atmospheric refraction across the small ($10\arcmin\times10\arcmin$) field of view of NIRC2, which primarily manifests as a very small change in the linear terms of the pixel-to-sky transformation that is typically negligible except for very high-precision astrometry.

Table~\ref{tbl:keck} reports all of the binary parameters we measured from our Keck/NIRC2 imaging and masking data. We present a total of 170 unique astrometric data sets here, 41 of which are from masking data. All binary parameters were first derived on individual images, and we computed the mean and rms from these. The rms was our initial estimate of the measurement error. We used this instead of the standard error on the mean because systematics like imperfect PSF modelling are not expected to vary independently from one image to another in a single data set. Each NIRC2 calibration also provides systematic uncertainties in their pixel scale and orientation correction. We adopted the larger quoted systematic for each of these parameters, corresponding to a fractional error in the pixel scale of $4.0\times10^{-4}$ and an error of 0.02\degree\ in PA. We added these systematics in quadrature to the rms of individual measurements to derive the final measurement error given in Table~\ref{tbl:keck}.

There is one more potential source of systematic error in our measurements, which is the uncertainty in the distortion solution. However, unlike the linear terms, there is no evidence that the distortion in NIRC2 varies from epoch to epoch at a level greater than the quoted 1.1\,mas error in the solution (as measured relative to an external astrometric catalogue). In order to exploit this, our observation strategy has been to acquire our targets at the same pixel location as often as possible from epoch to epoch. Thus, any systematic error in the distortion offsets applied to our astrometry should cancel out. In addition, many of our binaries are tight enough (separated by $\lesssim$10\,pixel) that the distortion is expected to be correlated for the two components, because distortion tends to vary over significantly larger spatial scales ($\sim$100\,pixel). For both of these reasons, it is likely that distortion contributes negligibly to our astrometric errors. It would be most significant for the widest binaries, but these are also the systems where the errors in the linear terms are largest. For example, a $\sim$2\arcsec\ ($\sim$200\,pixel) binary already has systematic errors of $\sim$1\,mas due to the uncertainty in the pixel scale and orientation of NIRC2. Therefore, we neglected the uncertainty in the distortion correction in our reported astrometric errors. 

For our purposes, the errors must accurately account for the epoch-to-epoch variations in our \emph{relative} astrometry. For other applications, such as combining our measurements with astrometry obtained with other instruments, it may be necessary to consider this additional systematic error. A handful of our binaries have resolved astrometry reported in \Gaia~EDR3. We choose not to include these measurements in our analysis until a better understanding of systematics in \Gaia\ measurements of close binaries, and the relationship between the \Gaia\ and NIRC2 astrometric reference frames, are available. For example, \citet{2021MNRAS.506.2269E} have recently shown that \Gaia's parallax uncertainties can be underestimated by up to a factor of three for binaries with separations $<$2\arcsec. Likewise, proper motions of such binaries have been known to harbour systematic errors since DR1 \citep{2017ApJ...840L...1M}. In our data, we also see evidence for a systematic offset between the PA of the NIRC2 frame and \Gaia, with our NIRC2 PA measurements being $0\fdg07\pm0\fdg02$ smaller on average than the PA reported by Gaia.

Finally, we note that we have previously published astrometry for one of the objects in our sample, KOI-3158AB a.k.a.\ Kepler-444AB, in \citet{2016ApJ...817...80D}. In that work, all of our astrometry was obtained before the AO realignment. We now have data taken after the realignment, and we have reanalysed all of our available data for this system using the appropriate astrometric calibrations. The first three epochs in 2013 and 2014 were taken at three different NIRC2 orientations and pixel locations of the target, so we have been obtaining data in each of those setups at later epochs in the hope of cancelling out distortion errors and combining all epochs into one self-consistent orbit fitting analysis. Unfortunately, the AO realignment in 2015 makes this impossible, so we have examined our post-realignment data sets in each of the three setups separately. We found consistent results between them, but we report the results only from the setup with the longest post-realignment time baseline (taken with the same pixel locations as our previously reported 2014~Nov~30~UT data).

\begin{table*}
\caption[]{Linear fits to the relative astrometry time-series measurements of planet-hosting binaries from Table~\ref{tbl:keck}.} \label{tbl:linfit}
\begin{tabular}{lcccccc}
\hline
System & $t_0$ & $\Delta{t}$ & $\rho_0$ & $\theta_0$ & $\dot{\rho}$ & $\dot{\theta}\rho_0$ \\
 & (MJD) & (yr) & (mas) & ($\degree$) & (\masyr) & (\masyr) \\ \hline
KOI-0001AB     & 56956 & 3.95 & $1109.37\pm0.27$\phn\phn & $136.368\pm0.012$\phn\phn     & \phs$   1.90\pm0.16$ & \phs$   0.15\pm0.14$     \\
KOI-0042AB     & 56927 & 4.11 & $ 1665.8\pm 0.4$\phn\phn & $ 35.581\pm0.012$\phn         &     $  -0.66\pm0.24$ & \phs$   0.52\pm0.20$     \\
KOI-0112AB     & 57405 & 2.93 & $  100.7\pm 0.5$\phn     & $ 114.13\pm 0.21$\phn\phn     & \phs$    0.0\pm 0.4$ &     $   -1.7\pm0.3$      \\
KOI-0214AB     & 57053 & 0.94 & $   68.0\pm 1.6$         & $  195.4\pm  1.5$\phn\phn     &     $     -6\pm   3$ &     $     -2\pm   4$     \\
KOI-0227AB     & 56985 & 2.96 & $ 299.73\pm0.11$\phn     & $ 69.002\pm0.019$\phn         & \phs$   0.01\pm0.09$ & \phs$   1.04\pm0.07$     \\
KOI-0249AB     & 56688 & 2.94 & $ 4333.0\pm 1.3$\phn\phn & $ 28.078\pm0.017$\phn         & \phs$    0.5\pm 0.9$ &     $   -0.4\pm 0.8$     \\
KOI-0270AB     & 57301 & 1.96 & $ 168.31\pm0.10$\phn     & $ 64.344\pm0.026$\phn         & \phs$   2.60\pm0.12$ & \phs$   0.38\pm0.11$     \\
KOI-0291AB$^*$ & 57349 & 2.90 & $  65.69\pm0.18$         & $  316.2\pm  0.4$\phn\phn     &     $  -0.47\pm0.16$ &     $   -0.3\pm 0.3$     \\
KOI-0588AB     & 56600 & 2.95 & $ 280.31\pm0.10$\phn     & $276.327\pm0.015$\phn\phn     & \phs$   0.44\pm0.07$ &     $  -0.93\pm0.06$     \\
KOI-0854AB     & 57071 & 3.18 & $   17.7\pm 0.6$         & $  222.1\pm  2.7$\phn\phn     & \phs$    1.0\pm 0.4$ & \phs$    2.6\pm 0.5$     \\
KOI-0975AB     & 57254 & 2.28 & $ 776.31\pm0.30$\phn     & $129.474\pm0.020$\phn\phn     & \phs$    3.3\pm 0.3$ & \phs$   1.21\pm0.29$     \\
KOI-1422AB     & 56863 & 2.03 & $ 214.16\pm0.11$\phn     & $216.927\pm0.023$\phn\phn     &     $  -0.76\pm0.14$ &     $  -0.33\pm0.10$     \\
KOI-1613AB     & 57225 & 4.90 & $ 207.60\pm0.05$\phn     & $184.519\pm0.016$\phn\phn     &     $  -1.47\pm0.04$ & \phs$   0.01\pm0.04$     \\
KOI-1615AB     & 56906 & 4.21 & $   24.8\pm 0.8$         & $  135.6\pm  1.5$\phn\phn     &     $   -3.5\pm 0.4$ & \phs$    2.7\pm 0.4$     \\
KOI-1619AB     & 57015 & 4.34 & $ 2068.8\pm 0.6$\phn\phn & $226.605\pm0.012$\phn\phn     & \phs$    1.0\pm 0.3$ &     $  -0.66\pm0.23$     \\
KOI-1681AB     & 56863 & 2.03 & $ 148.61\pm0.16$\phn     & $ 141.19\pm 0.03$\phn\phn     &     $  -1.83\pm0.22$ &     $  -0.13\pm0.09$     \\
KOI-1725AB     & 56974 & 0.82 & $ 4053.4\pm 1.2$\phn\phn & $ 98.567\pm0.013$\phn         &     $     -7\pm   4$ &     $   -2.5\pm 2.5$     \\
KOI-1835AB     & 57175 & 4.88 & $   53.9\pm 0.4$         & $ 355.49\pm 0.28$\phn\phn     &     $  -1.28\pm0.29$ & \phs$   0.45\pm0.18$     \\
KOI-1841AB     & 57048 & 4.90 & $ 309.46\pm0.12$\phn     & $ 74.097\pm0.020$\phn         & \phs$   0.90\pm0.06$ &     $  -0.24\pm0.05$     \\
KOI-1961AB$^*$ & 57432 & 2.92 & $  38.51\pm0.10$         & $  262.3\pm  0.7$\phn\phn     & \phs$   2.59\pm0.09$ & \phs$    2.0\pm 0.4$     \\
KOI-1962AB     & 57079 & 3.92 & $ 123.18\pm0.21$\phn     & $ 114.09\pm 0.05$\phn\phn     & \phs$   2.06\pm0.16$ & \phs$   0.30\pm0.09$     \\
KOI-1964AB     & 57064 & 4.02 & $ 386.18\pm0.09$\phn     & $  1.063\pm0.012$             &     $  -3.23\pm0.06$ &     $  -1.82\pm0.05$     \\
KOI-1977AB     & 56796 & 3.92 & $  81.08\pm0.15$         & $  77.03\pm 0.25$\phn         &     $  -1.45\pm0.10$ &     $   -1.2\pm 0.8$     \\
KOI-2005AB$^*$ & 57246 & 3.93 & $  20.26\pm0.30$         & $    237\pm    4$\phn\phn     & \phs$   0.89\pm0.27$ &     $   -1.1\pm 0.9$     \\
KOI-2031AB$^*$ & 57132 & 3.93 & $   56.0\pm 0.6$         & $  246.6\pm  1.4$\phn\phn     &     $   -0.3\pm 0.4$ &     $   -0.1\pm 0.9$     \\
KOI-2059AB     & 56511 & 1.96 & $ 385.44\pm0.21$\phn     & $289.573\pm0.020$\phn\phn     & \phs$   1.17\pm0.21$ &     $  -1.21\pm0.14$     \\
KOI-2124AB$^*$ & 56985 & 4.88 & $   56.5\pm 1.1$         & $  53.54\pm 0.15$\phn         & \phs$    3.1\pm 0.5$ & \phs$   0.04\pm0.07$     \\
KOI-2179AB     & 56861 & 2.02 & $  133.6\pm 0.4$\phn     & $ 356.09\pm 0.10$\phn\phn     &     $   -1.0\pm 0.4$ &     $  -0.10\pm0.22$     \\
KOI-2295AB     & 56966 & 3.94 & $ 2188.1\pm 1.0$\phn\phn & $ 78.563\pm0.014$\phn         &     $   -0.3\pm 0.7$ & \phs$    1.5\pm 0.3$     \\
KOI-2418AB     & 57345 & 2.94 & $  103.7\pm 0.6$\phn     & $    2.1\pm  0.3$             &     $   -1.5\pm 0.5$ & \phs$    0.1\pm 0.5$     \\
KOI-2542AB     & 56863 & 2.02 & $ 764.11\pm0.21$\phn     & $ 28.674\pm0.013$\phn         &     $  -0.26\pm0.26$ &     $  -0.42\pm0.21$     \\
KOI-2672AB     & 56869 & 1.97 & $ 640.09\pm0.23$\phn     & $306.502\pm0.020$\phn\phn     &     $  -2.23\pm0.25$ & \phs$   0.80\pm0.24$     \\
KOI-2705AB     & 57265 & 2.28 & $ 1888.4\pm 0.5$\phn\phn & $303.900\pm0.017$\phn\phn     &     $   -0.8\pm 0.5$ &     $   -0.8\pm 0.7$     \\
KOI-2733AB     & 57266 & 2.30 & $ 105.62\pm0.18$\phn     & $ 294.61\pm 0.13$\phn\phn     &     $  -2.47\pm0.22$ & \phs$   1.05\pm0.26$     \\
KOI-2790AB     & 57405 & 2.93 & $ 253.25\pm0.09$\phn     & $135.017\pm0.023$\phn\phn     &     $  -0.63\pm0.09$ & \phs$   0.36\pm0.10$     \\
KOI-3158AB     & 57325 & 0.99 & $ 1841.9\pm 0.4$\phn\phn & $252.834\pm0.013$\phn\phn     &     $   -0.6\pm 1.0$ &     $   -0.1\pm 0.9$     \\
KOI-3255AB     & 57266 & 2.30 & $ 181.21\pm0.07$\phn     & $336.006\pm0.027$\phn\phn     &     $  -0.42\pm0.09$ &     $  -0.15\pm0.10$     \\
KOI-3284AB     & 57046 & 0.98 & $ 438.69\pm0.17$\phn     & $192.802\pm0.020$\phn\phn     &     $   -0.5\pm 0.3$ &     $   -0.2\pm 0.3$     \\
KOI-3444AB     & 57222 & 1.84 & $1085.33\pm0.27$\phn\phn & $  9.718\pm0.014$             & \phs$    2.4\pm 0.4$ &     $   -0.6\pm 0.3$     \\
KOI-3892AB     & 57053 & 0.94 & $  120.8\pm 1.6$\phn     & $  341.8\pm  0.8$\phn\phn     & \phs$      8\pm   3$ & \phs$     10\pm   3$\phn \\
KOI-3991AB     & 57267 & 2.29 & $ 201.17\pm0.24$\phn     & $ 111.60\pm 0.03$\phn\phn     &     $  -1.63\pm0.21$ & \phs$   0.63\pm0.23$     \\
KOI-4032AB     & 56690 & 0.98 & $  126.2\pm 0.7$\phn     & $   30.6\pm  0.3$\phn         & \phs$    1.8\pm 1.4$ &     $   -0.2\pm 1.5$     \\
KOI-4097AB     & 57209 & 3.91 & $  176.8\pm 0.6$\phn     & $  17.84\pm 0.14$\phn         &     $   -1.5\pm 0.3$ & \phs$   5.33\pm0.24$     \\
KOI-4184AB$^*$ & 57227 & 3.91 & $ 206.47\pm0.05$\phn     & $ 223.25\pm 0.05$\phn\phn     & \phs$   0.30\pm0.03$ &     $  -0.23\pm0.13$     \\
KOI-4252AB     & 57488 & 2.88 & $  48.11\pm0.15$         & $ 341.29\pm 0.15$\phn\phn     & \phs$   3.05\pm0.10$ &     $  -3.53\pm0.08$     \\
\hline
\end{tabular}
\smallskip
\begin{list}{}{}
\item[\sc Note---]Linear polynomials were fitted to separation ($\rho$) as a function of time and PA ($\theta$) as a function of time. The measurements used in the fits are from Table~\ref{tbl:keck}, and the mean epoch ($t_0$) was subtracted from the input times before performing the fit so that the zeroth-order coefficient would provide the best-fit separation and PA at the mean epoch ($\rho_0$ and $\theta_0$, respectively). We report the first-order coefficients for both separation ($\dot{\rho}$) and PA ($\dot{\theta}$) in units of \masyr\ for clarity, using $\rho_0$ to convert the fitted PA slope from \degree\,yr$^{-1}$ to \masyr.  Polynomial fits use the measurement errors from Table~\ref{tbl:keck} to determine the coefficient uncertainties except where noted.  \item[*---]Using measurement errors in these fits resulted in unreasonably high $\chi^2$ values, $p(\chi^2) < 0.03$, so we instead used unweighted fits in these cases and report coefficient uncertainties that are scaled by $\sqrt{\chi^2/(N_{\rm meas}-2)}$ (i.e., the rms scatter about the fit is used to estimate the measurement errors).
\end{list}
\end{table*}

\section{Measuring Orbital Arcs \label{sec:arcs}}

It is expected that the binaries in our sample should generally display orbital motion that is indistinguishable from linear motion within our astrometric errors. The median estimated orbital period for our sample is 320\,yr, so even our longest observational time baseline of 5\,yr typically covers $\lesssim$2\% of the orbit. Among the sample of 45 binaries for which we present astrometry, the smallest estimated orbital period is 11\,yr for KOI-0854AB. We only have two epochs of astrometry for this system, so it is impossible to assess its nonlinear motion. The next shortest estimated period is 16\,yr for KOI-4252AB, and our data spanning 2.88\,yr do not show evidence of nonlinear motion. This provided a justification for assuming that linear motion was the appropriate model to fit to our astrometry.

We fitted each set of separations and PAs as a functions of time, yielding an instantaneous measurement of the orbital motion at the mean epoch of the input data. In our fits, we first subtracted the mean epoch ($t_0$) from the input times, so that the zeroth order coefficient in each fit would give the separation and PA and its uncertainty at the mean epoch ($\rho_0$ and $\theta_0$, respectively). We used the IDL routine \texttt{POLY\_FIT} to perform the fits and provided the astrometric errors as inputs in order to compute uncertainties in the fitted parameters and the $\chi^2$ of the best-fit solution. We also performed fits with no input errors in which parameter uncertainties were adjusted by the factor $\sqrt{\chi^2/(N-2)}$, where $N$ was the number of data points in the fit, thereby forcing the reduced $\chi^2$ to be $\approx1$. 

\begin{figure}
  \centerline{\includegraphics[width=0.45\textwidth]{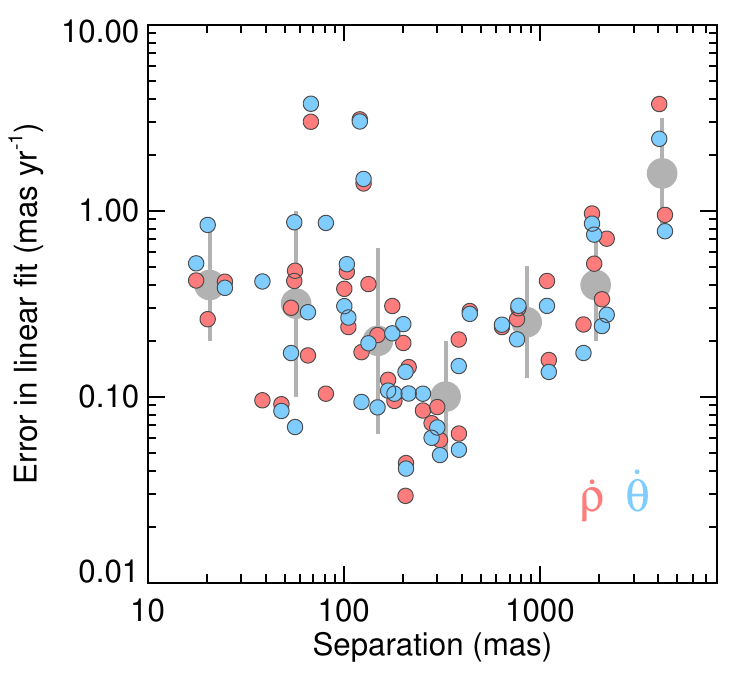}}
  \caption{\normalsize Uncertainties in the slopes computed from our linear fits to separation ($\dot{\rho}$, red) and PA ($\dot{\theta}$, blue) plotted as a function of binary separation. A major factor in these uncertainties is the per-epoch astrometric uncertainty, but they also reflect the range in time baselines and number of epochs in our orbit monitoring sample, which explains the scatter. Starting at the tightest binaries and going wider, astrometric errors improve (on average) as binaries become more cleanly resolved. Eventually uncertainties in the astrometric calibration of NIRC2 become significant at scales $\gtrsim500$\,mas, causing the average error to increase for the widest binaries. Large grey symbols show the average and rms of the individually plotted measurements divided into logarithmic bins that are a factor of 2.5 (0.4\,dex) wide. 
  \label{fig:arcerr}}
\end{figure}

In the vast majority of cases, we found reasonable $\chi^2$ values, and an ensemble mean $p(\chi^2) \approx 0.5$, but six of the 90 total fits had $p(\chi^2) < 0.03$. In principle, these could be indicative of nonlinear motion, but the residuals are not obviously consistent with Keplerian motion, so we conservatively assume that these are cases where our measurement errors are somewhat underestimated.  In these six cases we used the unweighted fits where the parameter uncertainties were adjusted to correspond to reduced $\chi^2\approx1$.

Table~\ref{tbl:linfit} gives a summary of our orbital-arc fitting results. The slopes are given in units of \masyr, using $\rho_0$ to convert our fitted $\dot{\theta}$ values from \degree\,yr$^{-1}$ to \masyr\ to be more easily comparable to the $\dot{\rho}$ values. Figure~\ref{fig:arcerr} shows how the uncertainties in our linear fits depends on binary separation. 

\begin{figure}
  \centerline{
    \hskip -0.03\textwidth
    \includegraphics[width=0.45\textwidth]{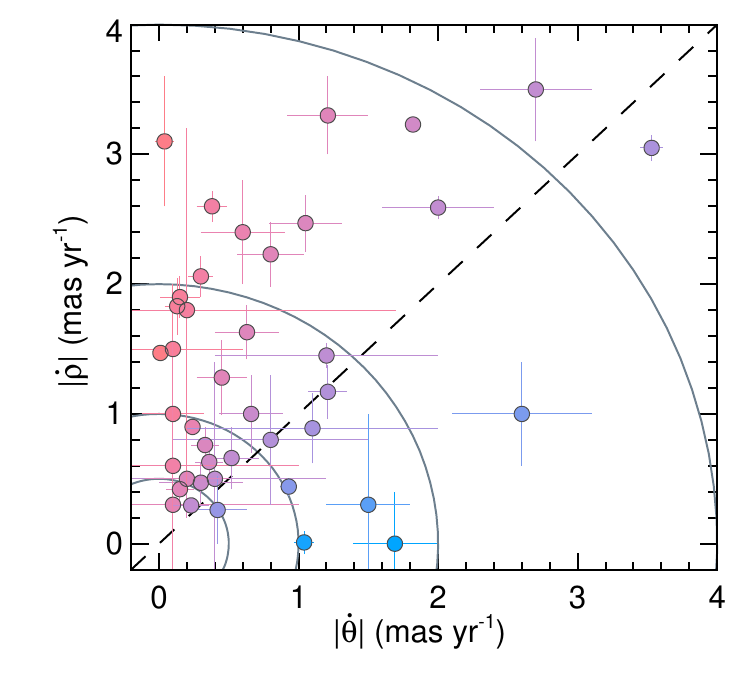}
    \hskip -0.40\textwidth
    \includegraphics[width=0.45\textwidth]{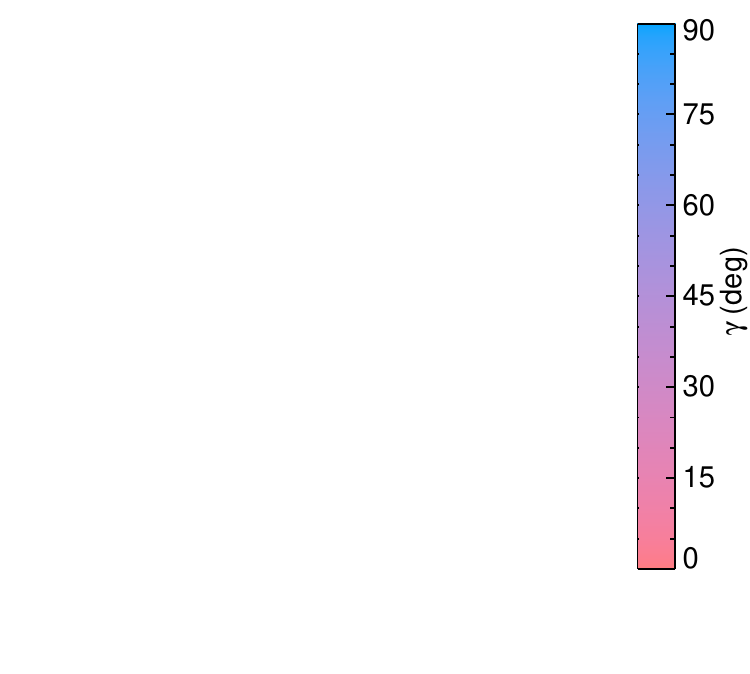}
    }
  \caption{\normalsize Amplitude of the orbital motion in the separation direction $|\dot{\rho}|$ as a function of the amplitude of orbital motion in the PA direction $|\dot{\theta}|$ for our sample. Colour indicates how consistent with being edge on (red) or face on (blue) the orbital motion is, as defined by the angle $\gamma$ (Equation~\ref{eq:gamma}). The dashed line is the line of equality ($|\dot{\rho}| = |\dot{\theta}|$), and the solid grey curves indicate total motion of 0.5, 1, 2, and 4\,\masyr. \label{fig:vt-vr}}
\end{figure}

\subsection{Confirming physical association \label{sec:bound}}

Most of the binaries in our sample have not been observationally confirmed as being physically bound, common proper motion pairs, although the simulation of background stars for this sample by \citet{2016AJ....152....8K} strongly implies that all but the widest, faintest companions are very likely to be physical. We tested whether any of our sample might be unassociated pairs by using the criterion $B$ defined by \citet{2015MNRAS.448.3679P} as
\begin{equation}
\hfill
B \equiv \frac{1}{8\pi^2}\left(\frac{d}{{\rm pc}}\right)^3\left(\frac{\sqrt{\dot{\rho}^2+\dot{\theta}^2}}{{\rm arcsec / yr}}\right)^2\left(\frac{\rho_0}{{\rm arcsec}}\right)\left(\frac{\Mtot}{ M_\odot}\right)^{-1}.
\label{eq:B}
\hfill
\end{equation}
This criterion depends on the distance ($d$), total angular velocity, angular separation, and total mass of the system (\Mtot). For bound companions, $B<1$. We used the same distances and masses as in Section~\ref{sec:sample} to compute $B$ and assumed a 10\% uncertainty in \Mtot\ when propagating the error in $B$. 

All but three systems are clearly bound (having $B<1$ at 3$\sigma$ or higher). These exceptions are KOI-0214AB (1.1$\sigma$ above $B=1$), KOI-1725AB (0.3$\sigma$ below $B=1$), and KOI-3892AB (2.0$\sigma$ above $B=1$). KOI-1725AB is one of the few systems to have already been confirmed as a common proper motion (and common parallax) pair. \citet{2017AJ....153..267M} measured resolved astrometry for this 4\arcsec\ binary using CFHT/WIRCam seeing-limited imaging, and this is also confirmed in \Gaia~EDR3 \citep{2021_GaiaEDR3_Summary}. 

All three of these systems are notable for having the largest uncertainties in their angular velocities, $\gtrsim$10$\times$ higher than the median 0.25\,\masyr. KOI-0214AB and KOI-3892AB have such poor astrometry because they are exceptionally tight and faint ($\rho_0 = 70$\,mas and 120\,mas and $\Delta\Kp\approx4$\,mag), which makes high-precision astrometry very challenging to obtain. If their companions were zero-proper-motion background sources, they would have appeared to move 6\,\masyr\ and 37\,\masyr, respectively, which would have been detectable at 2$\sigma$ and 12$\sigma$. We determine that it is most likely that these noisiest angular velocities resulted in $B$ values that scattered high, rather than the alternative that the only unbound systems in the sample happened to be those with the poorest astrometry. The latter scenario is very unlikely {\it a priori}, as any background contaminants in our sample are expected to be at $\gtrsim$2$\times$ wider separations (see Figure~4 of \citealp{2016AJ....152....8K}).

We conclude that all of the systems in our sample are physically bound binaries. We note that the lack of any apparently unbound systems in our sample is consistent with the expectation that unresolved higher-order multiplicity should be negligible in our sample. Such inner binaries could systematically alter the observed linear motion of the photocenter motion, which would be unlikely to preserve $B<1$.

\section{Tests of Orbital Alignment \label{sec:test}}

For an ensemble of binary systems that each host one or more transiting planets, perhaps the simplest orbital test that can be performed is whether the stellar and planetary orbital planes are aligned. This is because all of the planets are known to be in edge-on orbits, to within a few degrees, and astrometry provides a straightforward test of whether the binary orbits are also edge on. For any given binary, if there is significant motion in PA, then the orbit is not edge on. In practice, however, we are not interested in testing for mutual inclinations of exactly zero, as even our relatively flat solar system has misaligned orbits up to a few degrees. In addition, observing a preference for more-aligned-than-not orbits, rather than extremely flat, would still be interesting astrophysically. 

Alignment tests on single systems can never be definitive. In our case, the PA of the transit chord is not known, and so both the transit and binary can been seen edge on even if their angular momentum vectors are highly misaligned in a direction that happens to be orthogonal to the plane of the sky. Therefore, a statistical sample is necessary to perform any robust alignment tests.

In the following, we examine our orbital measurements in a variety of ways to determine which underlying distributions of mutual inclination between the stellar and planetary orbital planes are most likely, and which are ruled out. For simplicity, we will generally refer to low mutual inclinations as ``aligned'' and higher mutual inclinations as ``misaligned.''

\subsection{A binary variable test \label{sec:binvar}}

We first approach the question of alignment by turning our measurements into a binary variable, that is, either true or false. This approach has the advantages of being intuitively simple and amenable to statistical tests like the binomial theorem. The binary variable we consider is whether the amplitude of the motion in the separation direction $\abs{\dot{\rho}}$ is larger or smaller than the motion in the PA direction $\abs{\dot{\theta}}$. We refer to these two cases as more-aligned or less-aligned, respectively.

We performed a Monte Carlo simulation of the null hypothesis that binary orbital inclinations are distributed with isotropic viewing angles, $p(i) \propto \cos{i}$. We assumed a uniform eccentricity distribution with $0.1 < e < 0.8$ and random times of observations. We found that 35\% of the trials resulted in more-aligned orbital arcs where $\abs{\dot{\rho}} > \abs{\dot{\theta}}$. This may seem unexpected, as the isotropic viewing angle prior means that edge-on orbits are more common, and more edge-on orbits have lower $\abs{\dot{\theta}}$. However, $\abs{\dot{\rho}}$ does not necessarily go up when $\abs{\dot{\theta}}$ goes down, as the orbital velocity also gets projected orthogonal to the plane of the sky. For example, at the median inclination of $i=60\degree$, a circular orbit maintains $\abs{\dot{\theta}} > \abs{\dot{\rho}}$ over the entire orbit.

Figure~\ref{fig:vt-vr} shows our orbital arc measurements. Out of the 45 binaries in our sample, 33 (73\%) have more-aligned orbital arcs, about two times higher that expected from the null hypothesis. Assuming that each binary represents an independent trial with a probability of 35\% of having a more-aligned orbital arc, the probability that 33 out of 45 systems would appear to be more aligned is $2\times10^{-7}$ according to the binomial theorem. Therefore, we conclude that the null hypothesis is ruled out at high significance corresponding to $\approx$5$\sigma$. We note that this result is independent of measurement uncertainties, as individual measurements are equally likely to scatter from more-aligned to less-aligned as the opposite.

This test relies on an assumption for the eccentricity distribution of the sample, and sufficiently high eccentricities might mimic more-aligned orbits. For instance, we performed another Monte Carlo test using a uniform distribution of $0.75 < e < 0.95$ that yielded a probability of 65\% for $\abs{\dot{\rho}} > \abs{\dot{\theta}}$. This modified version of the null hypothesis would be consistent with our result (15\% probability from the binomial theorem). This would be a remarkable result in its own right, as such an eccentricity distribution would be highly inconsistent with field binary statistics \citep[e.g.,][]{2010ApJS..190....1R}. To examine the influence of alignment versus eccentricity in more detail, we next consider tests using a continuous variable.

\begin{figure}
  \centerline{\includegraphics[width=0.5\textwidth,angle=0]{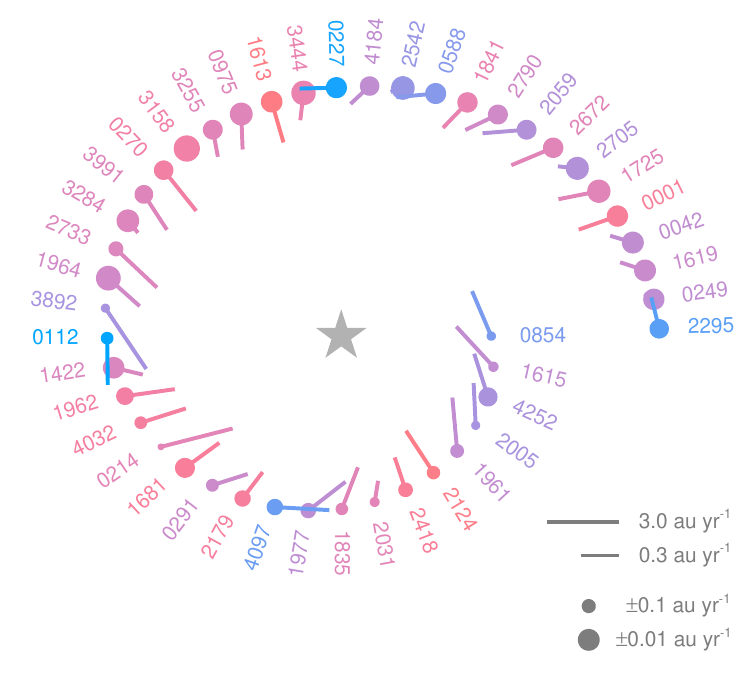}}
  \caption{\normalsize Pictorial representation of our orbital motion measurements. Binary companions are arranged in a spiral pattern, in order of increasing separation in AU (but not shown to scale), around a star symbol that depicts the (typically) planet-hosting primary. The line segments show the orbital motion direction (edge-on orbits point toward the central star), with longer segments indicating higher speeds. The size of the circle anchoring each line segment corresponds to the measurement precision (larger symbols have more precisely measured orbital velocities). Each system is labelled with its KOI number and colour-coded according to the orbital direction parameter $\gamma$ using the same colour mapping as in Figure~\ref{fig:vt-vr}. Orbital speeds are higher for more closely separated binaries, as expected for Keplerian motion, and wider binaries tend to have higher precision astrometry as they are easier to spatially resolve. More edge-on orbits (redder colour) are the most common, and more face-on orbits (bluer colour) appear at all binary separations.  \label{fig:spiral}}
\end{figure}

\begin{figure}
  \centerline{\includegraphics[width=0.5\textwidth,angle=0]{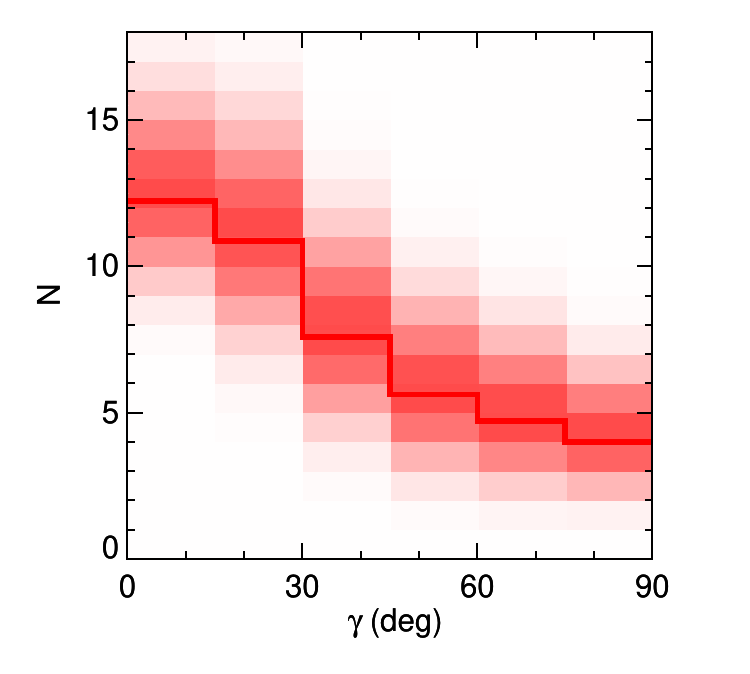}}
  \caption{\normalsize Histogram of the angle $\gamma$ that characterises the orbital direction of the binaries in our sample. The solid red line indicates an average over $10^4$ Monte Carlo trials with Gaussian noise added to each orbital motion measurement according to its error. The shading at different values of $N$ for each bin indicates the fraction of Monte Carlo trials resulting in that value of $N$ for that bin. The distribution peaks at $\gamma<15$\degree, consistent with a preponderance of nearly edge-on orbits. \label{fig:gamma-kois}}
\end{figure}

\subsection{A distance- and mass-independent variable test \label{sec:gamma}}

Our orbital arcs are essentially an instantaneous measure of the relative velocity between the two stars in the system. By definition, this velocity can be characterised as a direction and a speed. Of the two components of a velocity, the direction is an observable that depends on only on the orbital phase at the time of the observation and a few of the orbital parameters, namely $e$, $i$, and $\omega$. 

The orbital speed is a more complicated observable. First, it is measured in angular units, requiring a distance to convert into physical units. More important, deriving orbital speeds requires two more parameters, $a$ and \Mtot. Given the importance of accurate masses in analysing orbital speeds, we reserve this topic for a future study. Accurate stellar properties of even the single stars in the \Kepler\ sample are still being refined \citep[e.g.,][]{2020AJ....159..280B}, and commensurate work on the binary components is still ongoing (Sullivan et~al., in prep.; Ali et~al., in prep.). We also note that even though \Gaia\ has now reported parallaxes for our sample, most of them have large re-normalised unit weight errors (RUWE), implying that the \Gaia\ astrometric solutions harbour systematic errors caused by the binary nature of these sources.

In our present analysis we focus on directional information alone. Following the previous work on orbital arcs of wide binaries by \citet{1998AstL...24..178T}, we use the angle between the direction of orbital motion and the binary PA ($\gamma$) as the mass- and distance-independent variable.  We computed this angle from our orbital motion measurements, taking the absolute value of the motion in order for $\gamma$ to span only 0\degree--90\degree, as
\begin{equation}
    \hfill
    \gamma \equiv \arctan( \abs{\dot{\theta}}, \abs{\dot{\rho}} ).
    \hfill
    \label{eq:gamma}
\end{equation}
The sign of $\gamma$ is unimportant, as it simply corresponds to clockwise or counterclockwise motion. If the relative orbit appears to be changing only in separation and not in PA, as in an edge-on orbit, then $\gamma$ is 0\degree. If the motion is entirely in the PA direction, as in a face-on circular orbit, then $\gamma$ is 90\degree.

Figure~\ref{fig:spiral} summarises our orbital arc measurements, showing the speed, direction, and astrometric precision as a function of binary separation in AU. Figure~\ref{fig:gamma-kois} shows the distribution of $\gamma$ for the 45 orbit arcs we measured. In this histogram, as in all the following analysis, we account for the uncertainties in our orbital motions in a Monte Carlo fashion, propagating their Gaussian-distributed values into $\gamma$ values that do not necessarily follow Gaussian distributions.

We performed rank correlation tests of both speed and direction as a function of projected separation in AU. As expected for Keplerian orbits, we found that speed is correlated strongly with separation, with a Spearman's rank coefficient of $-0.56\pm0.04$ (probability of $4\times10^{-4}$ for the null hypothesis of no correlation). In contrast, we found no evidence of correlation of orbital direction with separation, with a Spearman's rank coefficient of $-0.005\pm0.083$. This is not necessarily expected, and in fact, it demonstrates that within our sample there is no evidence of any trend between planet-binary mutual inclination and binary separation.

\subsubsection{Simulating orbital arcs \label{sec:sims}}

To provide a comparison to our measured distribution of $\gamma$, we performed Monte Carlo simulations generating synthetic orbits that varied only the underlying distributions for inclination and eccentricity. In all simulations, the argument of periastron ($\omega$) and PA of the ascending node ($\Omega$) were drawn from uniform distributions from 0\degree--360\degree. The semimajor axis and period were fixed at unity, the time of periastron passage was fixed at zero, and the times of the synthetic observations were drawn from a uniform distribution between zero and unity. At each of these times, the separation and PA were computed as well as at two other times $\pm$0.01\% of the period before and after. The average of the difference in motion before and after the central time provided our simulated linear orbital motion.

We considered three forms for the binary inclination distribution. The first, simplest, represents isotropic viewing angles in which inclination is computed as $\arccos(\mathcal{U})$, where $\mathcal{U}$ is a uniformly distributed variate from zero to unity. 

The other distributions we tested represent a range in mutual inclination with respect to the planet orbit. Adopting similar notation to \citet{2019ApJ...883...22C}, the mutual inclination ($\phi$) between the planetary orbital plane and the stellar binary orbital plane can be expressed in terms of their orbital elements as
\begin{equation}
    \hfill
    \cos{\phi} = \cos{\ipl}\cos{\istar} + \sin{\ipl}\sin{\istar}\cos(\Omega_{\rm pl}-\Omega_{\star}).
    \hfill
    \label{eq:cosphi}
\end{equation}
Inverting this equation to solve for $\istar$ as a function of the other parameters gives a complicated set of equations, so instead we make the approximation that for transiting planets $\sin{\ipl} \approx 1$ and $\cos{\ipl} \approx 0$. The binary inclination can then be written simply as
\begin{equation}
    \hfill
    \istar = \arcsin\left(\frac{\cos{\phi}}{\cos{\Delta\Omega}}\right),
    \hfill
    \label{eq:istar}
\end{equation}
where we define  $\Delta\Omega \equiv \Omega_{\rm pl}-\Omega_{\star}$  as a relative quantity because we have no information on $\Omega_{\rm pl}$ for transiting planets. 

We generated randomly distributed values for $\istar$ by drawing random values for $\phi$ and $\Delta\Omega$. The two different types of mutual inclination distributions we considered had $\phi$ drawn uniformly between either $0\degree < \phi < \phimax$ (i.e., star-planet alignment within $\phimax$) or $\abs{\phi-\phi_0} < 5$\degree\ (i.e., misaligned by a narrowly specified amount). We assumed a uniform distribution from 0\degree--360\degree\ for the nuisance parameter $\Delta\Omega$ and ensured that $\abs{\cos{\Delta\Omega}} > \abs{\cos{\phi}}$, so that the argument of the $\arcsin$ in Equation~\ref{eq:istar} was valid. 

Overall, we tested ten different distributions for the mutual inclinations: isotropic; aligned within $\phi<\phimax$ for $\phimax = (10\degree,20\degree,30\degree,40\degree,50\degree)$; and misaligned within a $\pm5\degree$ band centred on $\phi_0 = (15\degree,25\degree,35\degree,45\degree)$. (We note that ``misalignment'' of $\phi_0=5$\degree\ is equivalent to aligned within $\phi_{\rm max}=10$\degree.) 

\begin{figure}
  \hskip 0.05in
  \centerline{
    \includegraphics[width=1.4in]{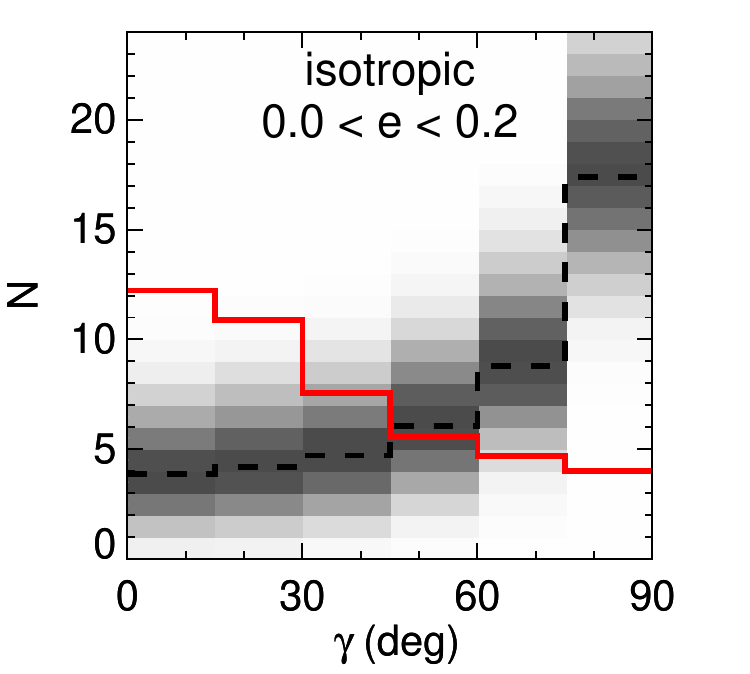}\hskip -0.35in
    \includegraphics[width=1.4in]{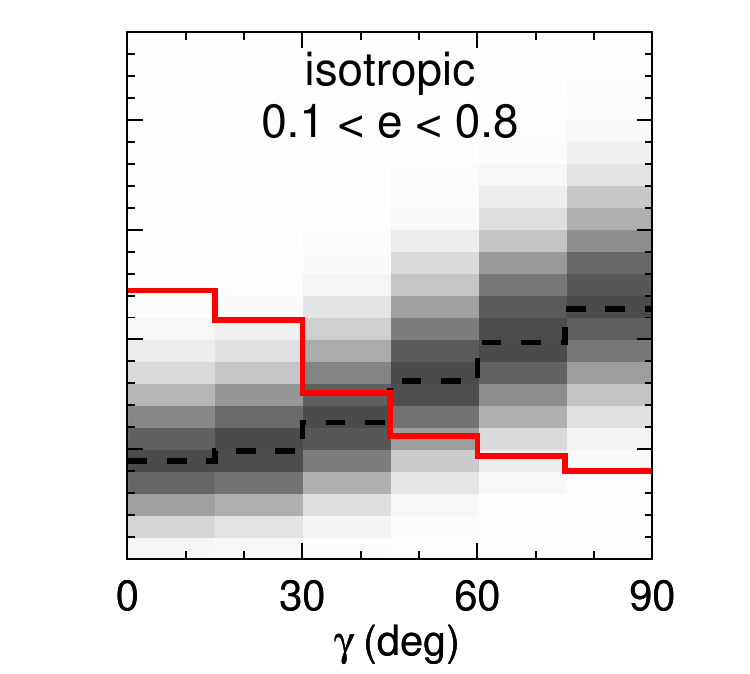}\hskip -0.35in
    \includegraphics[width=1.4in]{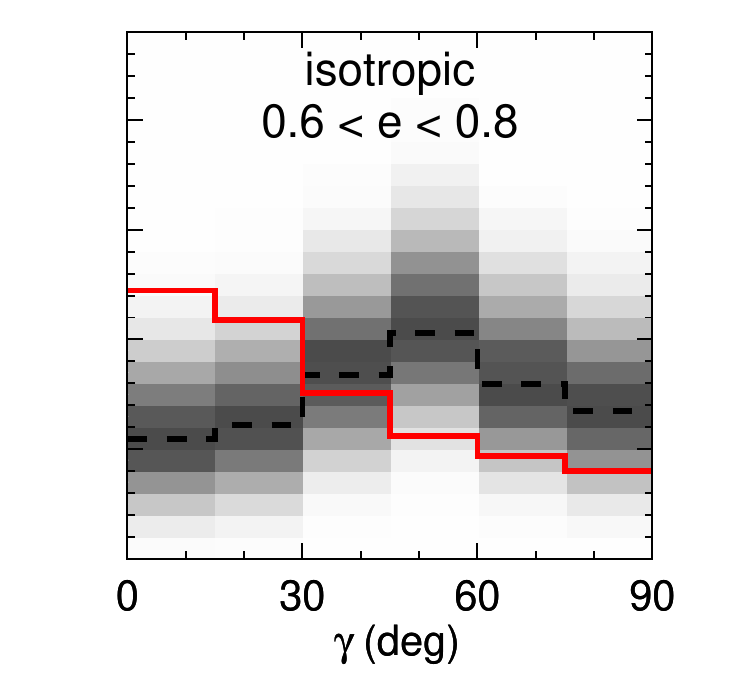}}\vskip -0.05in
  \caption{\normalsize Histograms of simulated $\gamma$ values for isotropic inclinations (i.e., uncorrelated binary and planet orbits) and different eccentricity distributions that range from more circular orbits (left) to field binary-like (middle) to highly eccentric (right). The dashed black line indicates an average over $10^5$ Monte Carlo trials, and the grey shading indicates sampling variance expected for a sample of 45 binaries not including measurement errors. The observed distribution from Figure~\ref{fig:gamma-kois} is over-plotted as a solid red line. Regardless of eccentricity, an isotropic distribution of binary orbital inclinations does not match our observations, implying that the binary orbits are not random and instead correlated with the planetary orbits.
  \label{fig:sim-hist-iso}}
\end{figure}

\begin{figure}
  \hskip 0.05in\centerline{
    \includegraphics[width=1.4in,angle=0]{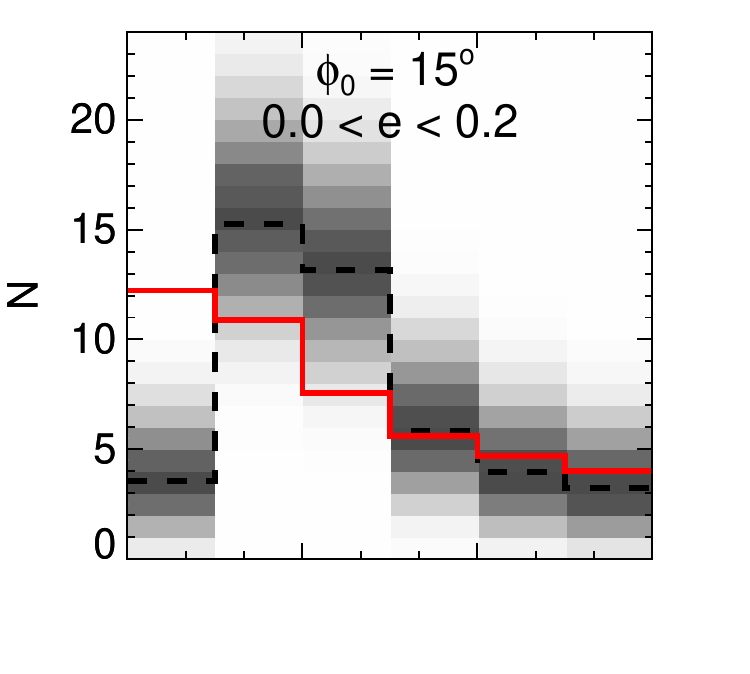}\hskip -0.35in
    \includegraphics[width=1.4in,angle=0]{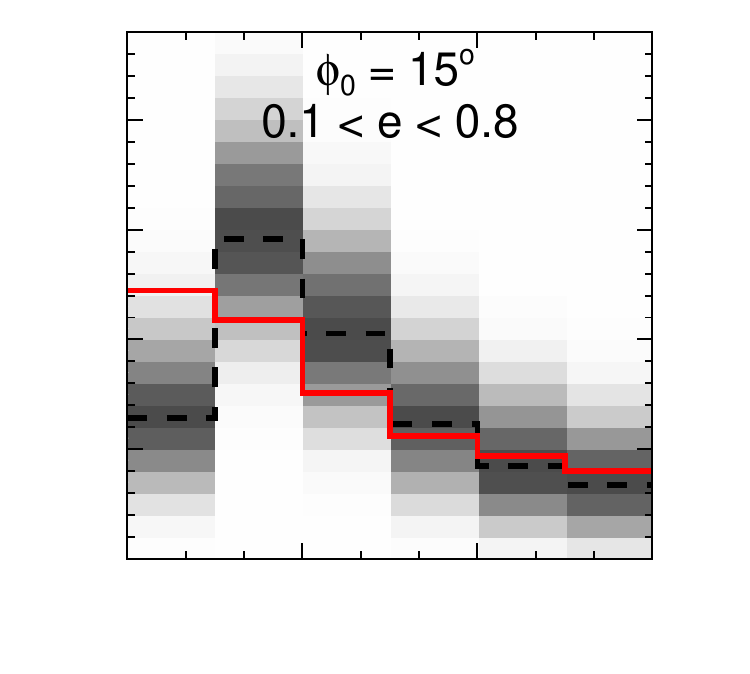}\hskip -0.35in
    \includegraphics[width=1.4in,angle=0]{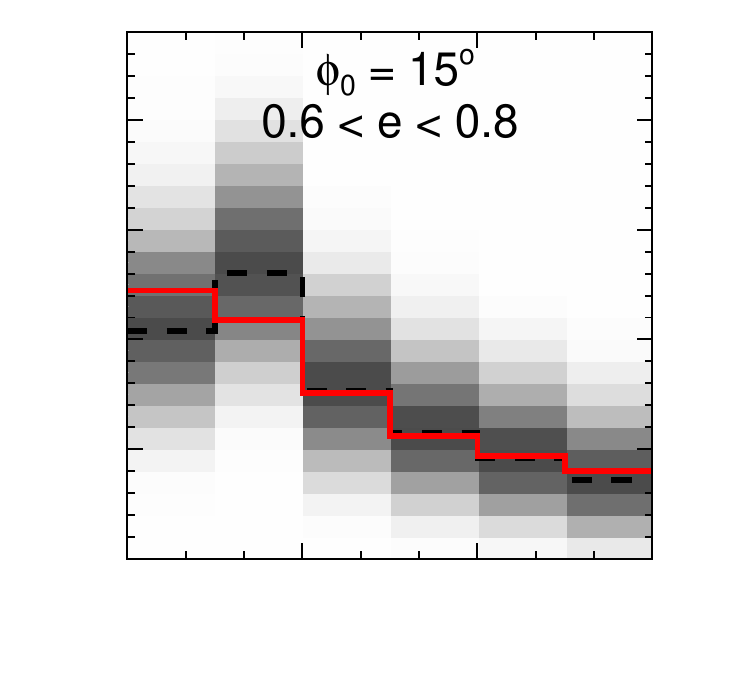}}\vskip -0.24in
  \hskip 0.05in\centerline{
    \includegraphics[width=1.4in,angle=0]{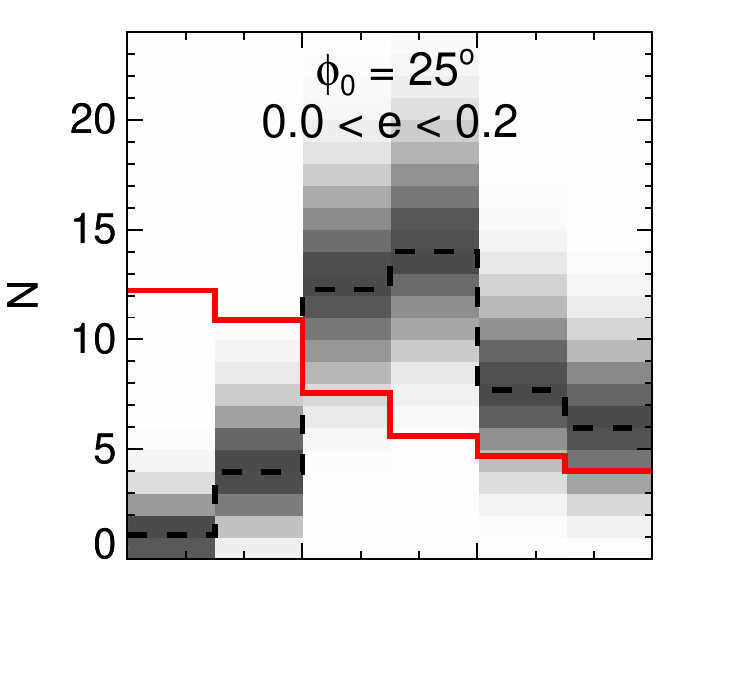}\hskip -0.35in
    \includegraphics[width=1.4in,angle=0]{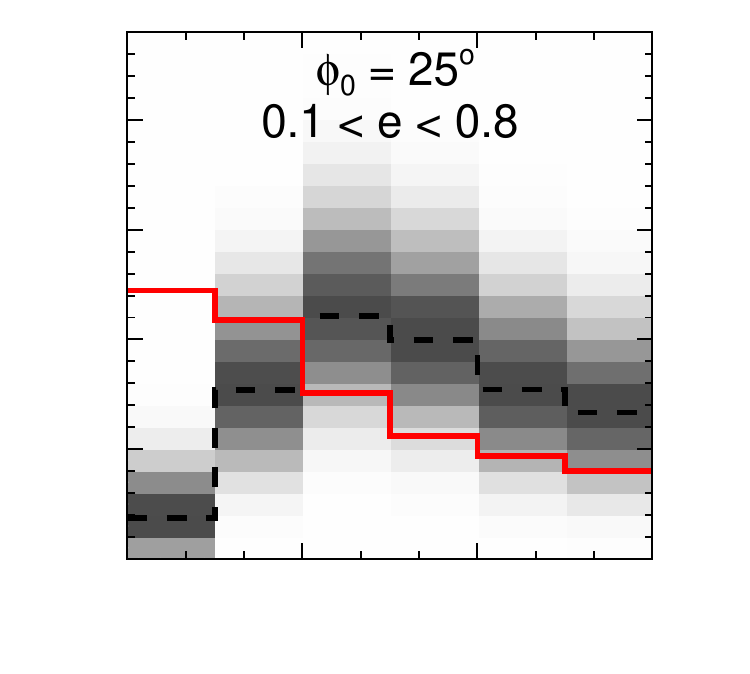}\hskip -0.35in
    \includegraphics[width=1.4in,angle=0]{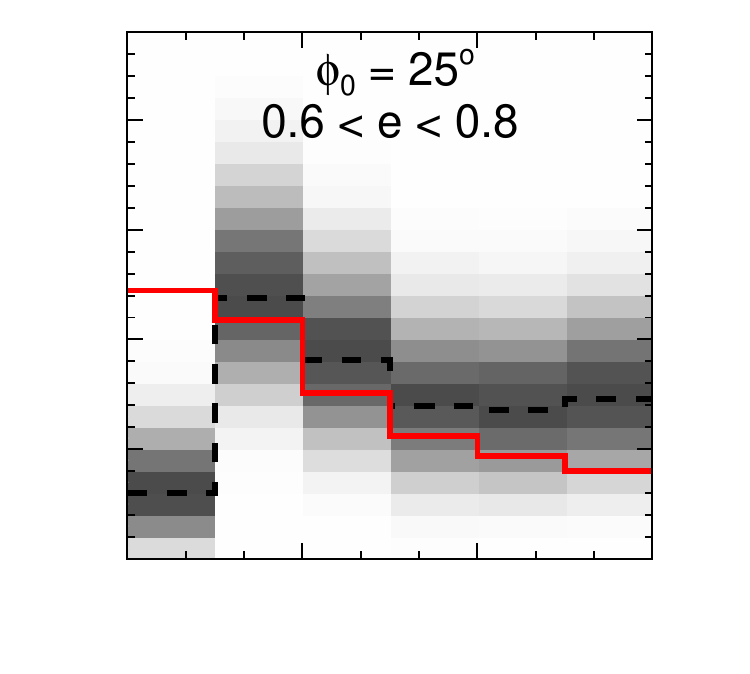}}\vskip -0.24in
  \hskip 0.05in\centerline{
    \includegraphics[width=1.4in,angle=0]{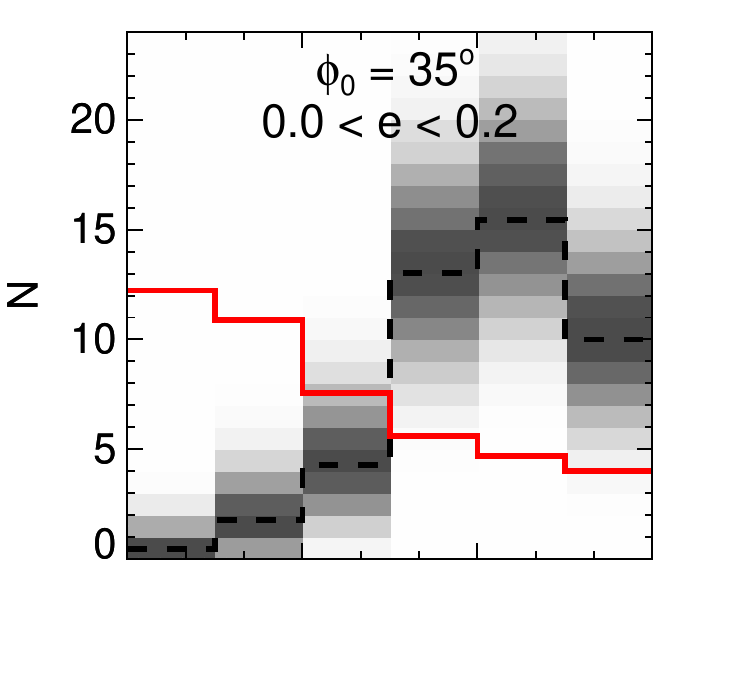}\hskip -0.35in
    \includegraphics[width=1.4in,angle=0]{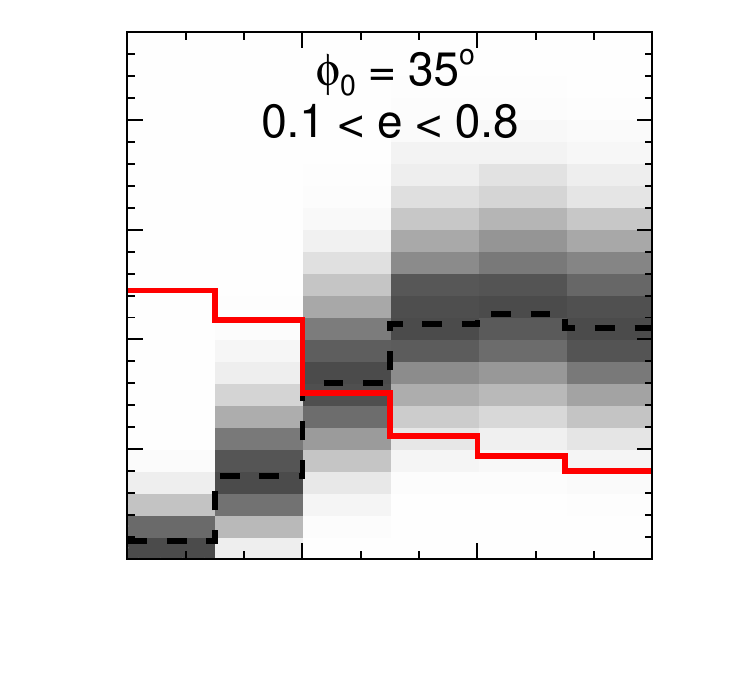}\hskip -0.35in
    \includegraphics[width=1.4in,angle=0]{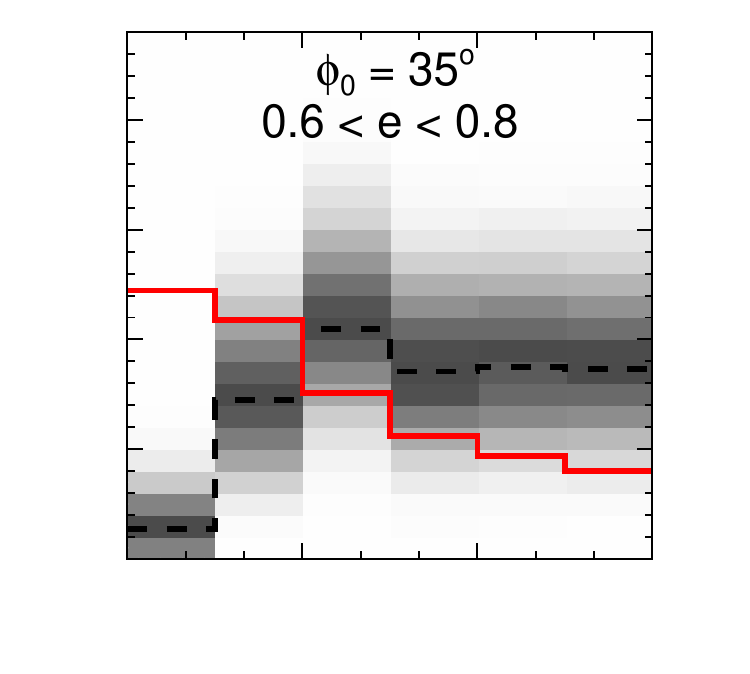}}\vskip -0.24in
  \hskip 0.05in\centerline{
    \includegraphics[width=1.4in,angle=0]{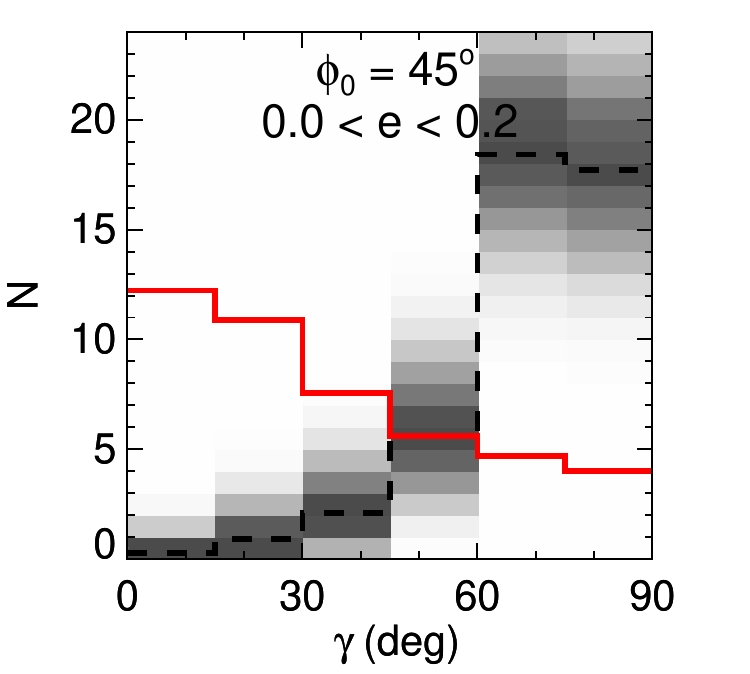}\hskip -0.35in
    \includegraphics[width=1.4in,angle=0]{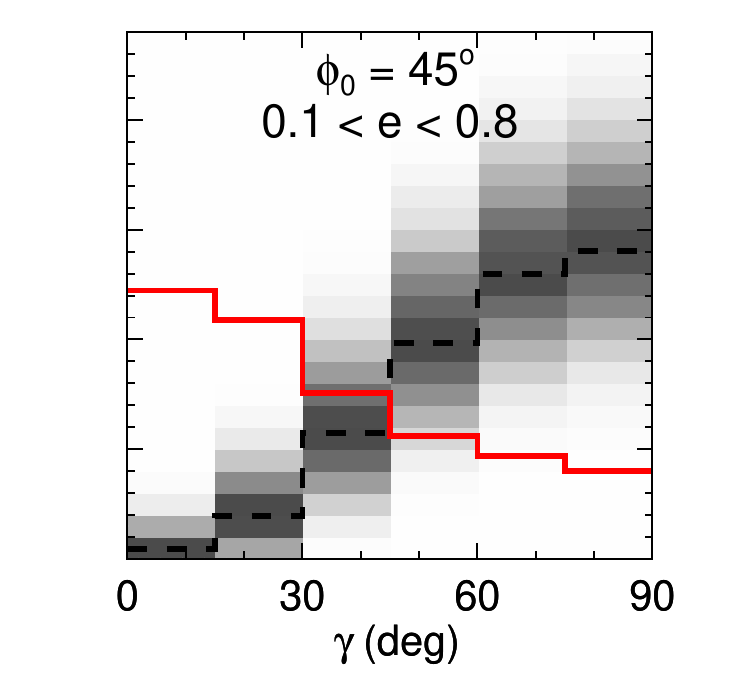}\hskip -0.35in
    \includegraphics[width=1.4in,angle=0]{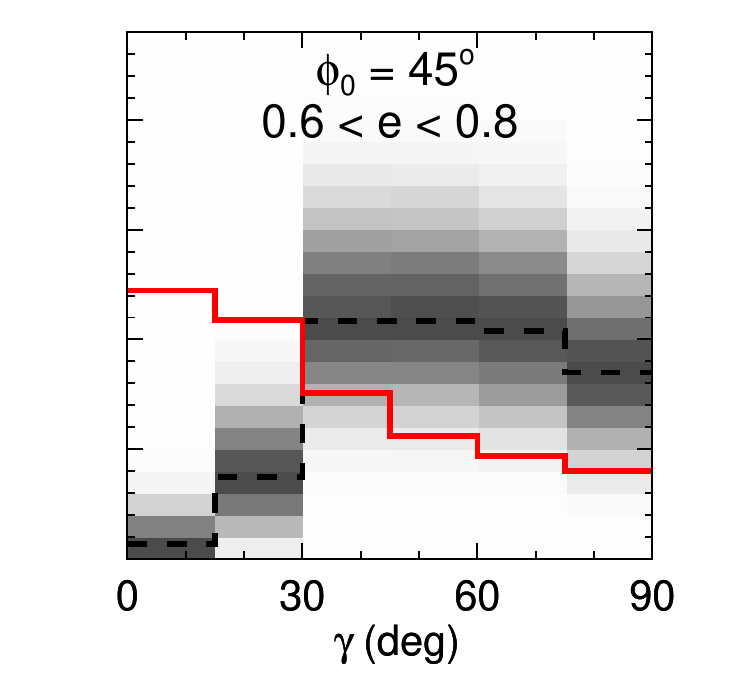}}\vskip -0.05in
  \caption{\normalsize Histograms of simulated $\gamma$ values, same as Figure~\ref{fig:sim-hist-iso}, except here the planet-binary mutual inclinations are misaligned by the angle $\phi_0 \pm 5\degree$. Misalignment increases from top to bottom, and the most misaligned cases are highly inconsistent with the observed $\gamma$ distribution, even for highly eccentric orbits. The bottom row ($\phi_0=45\degree$) corresponds to the mutual inclinations expected as the outcome of Kozai-Lidov planet migration.
  \label{fig:sim-hist-phi0}}
\end{figure}

\begin{figure}
  \hskip 0.05in\centerline{
    \includegraphics[width=1.4in,angle=0]{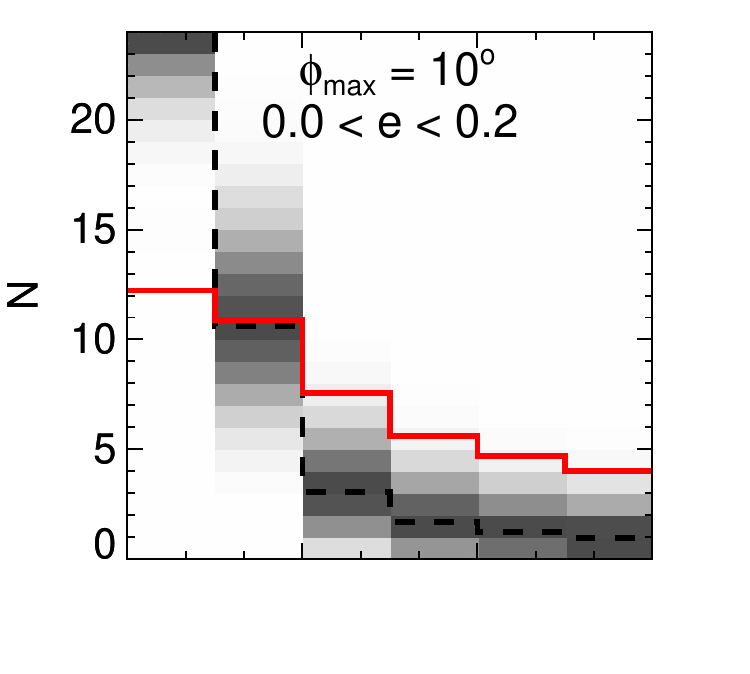}\hskip -0.35in
    \includegraphics[width=1.4in,angle=0]{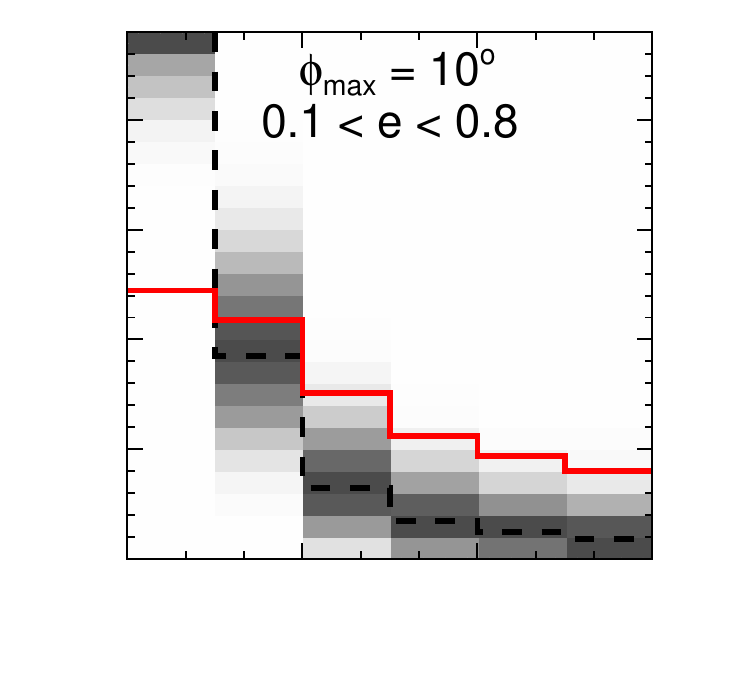}\hskip -0.35in
    \includegraphics[width=1.4in,angle=0]{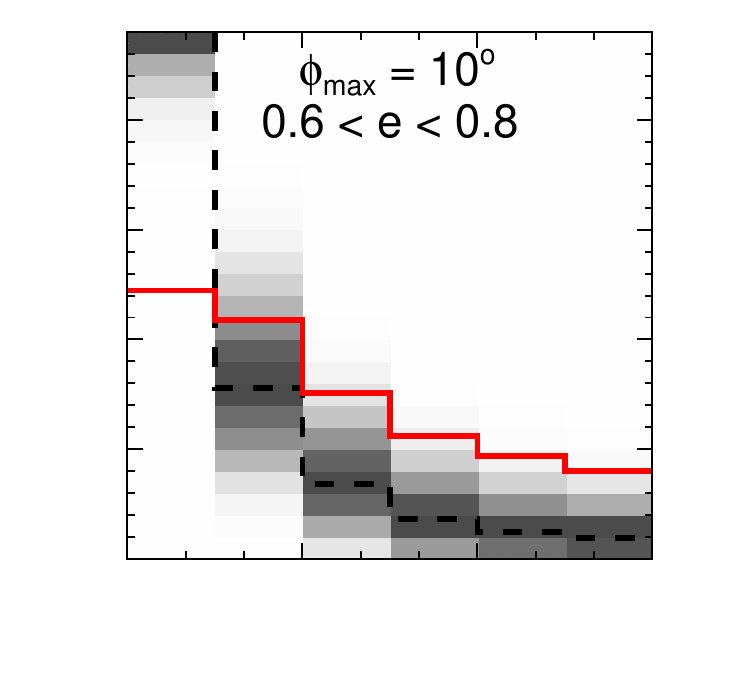}}\vskip -0.24in
  \hskip 0.05in\centerline{
    \includegraphics[width=1.4in,angle=0]{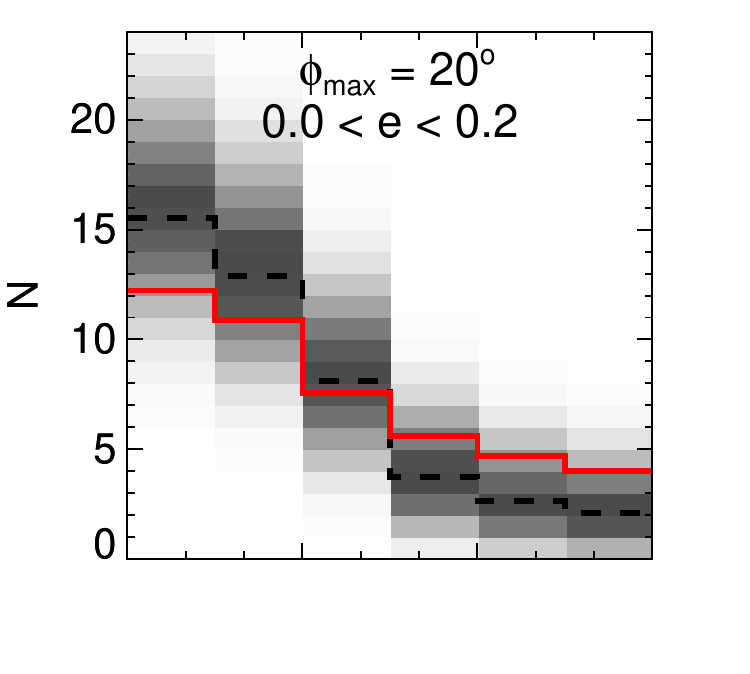}\hskip -0.35in
    \includegraphics[width=1.4in,angle=0]{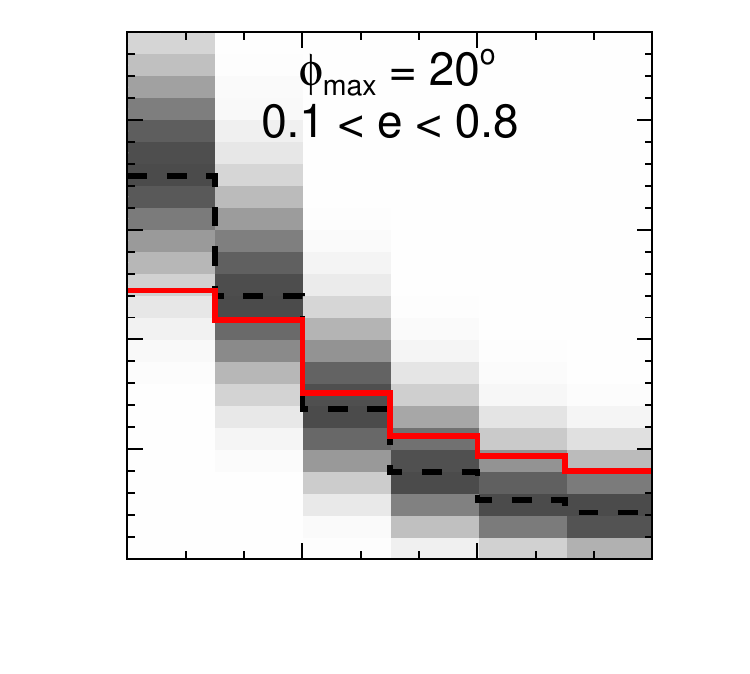}\hskip -0.35in
    \includegraphics[width=1.4in,angle=0]{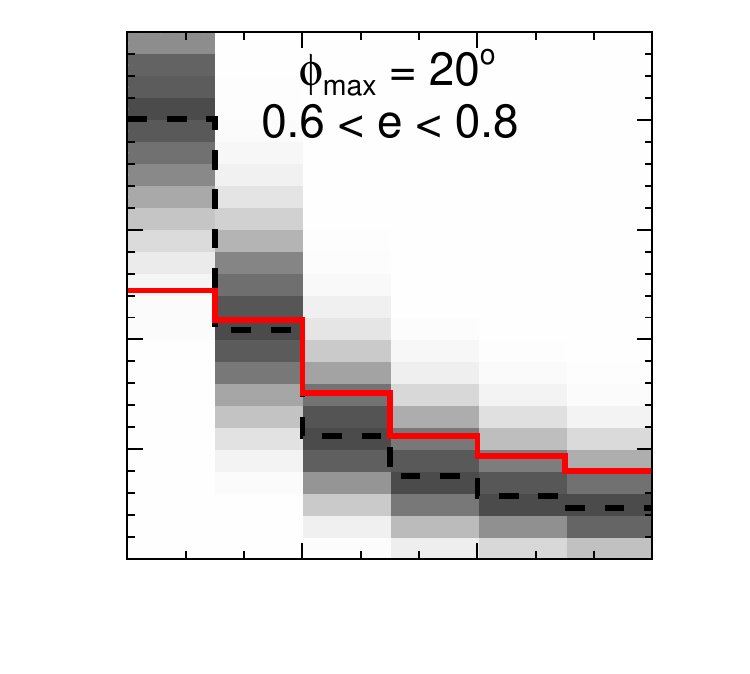}}\vskip -0.24in
  \hskip 0.05in\centerline{
    \includegraphics[width=1.4in,angle=0]{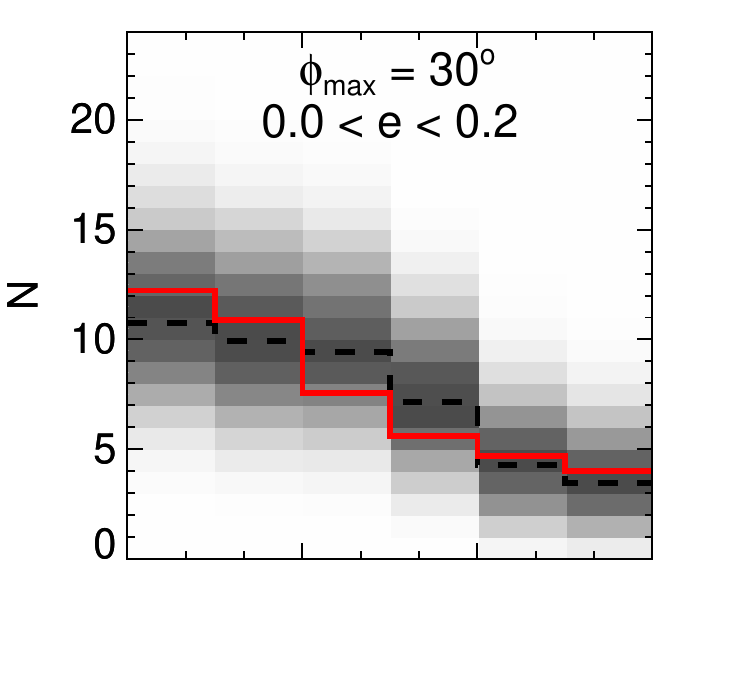}\hskip -0.35in
    \includegraphics[width=1.4in,angle=0]{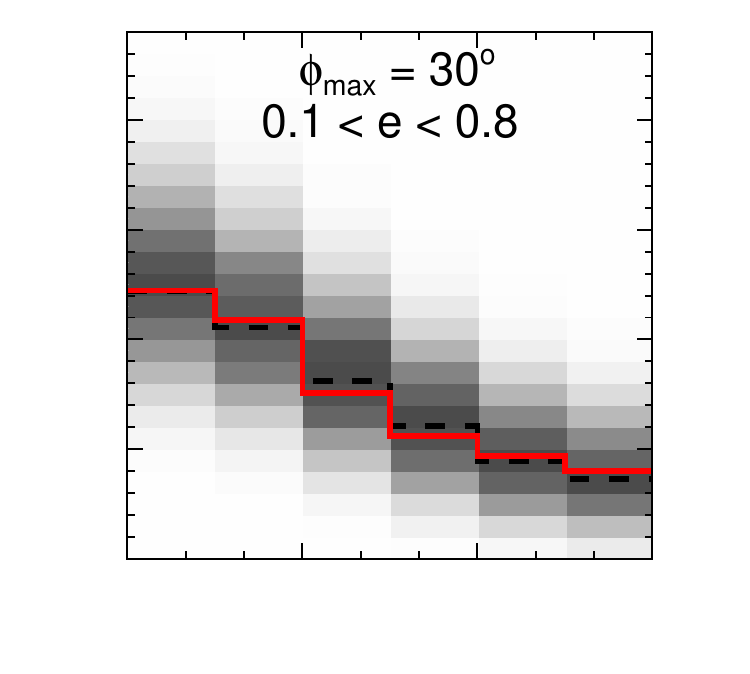}\hskip -0.35in
    \includegraphics[width=1.4in,angle=0]{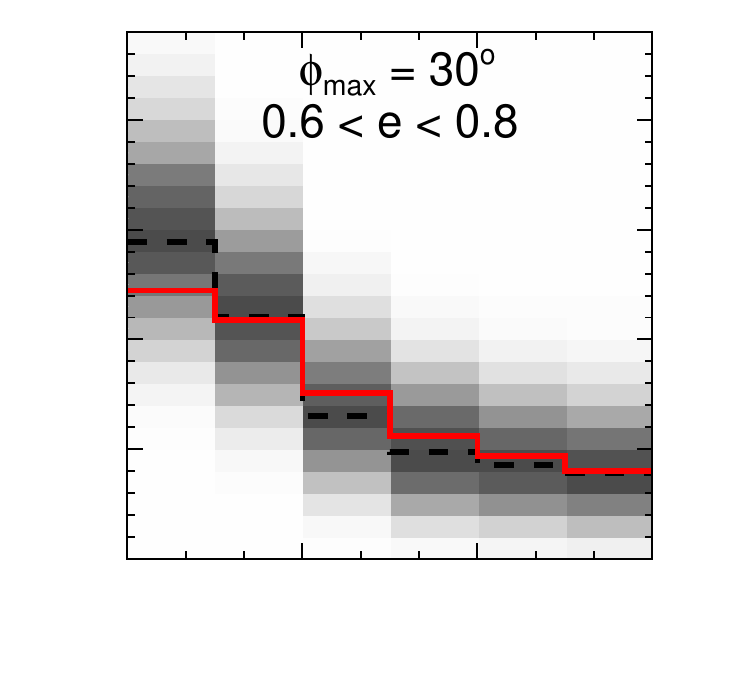}}\vskip -0.24in
  \hskip 0.05in\centerline{
    \includegraphics[width=1.4in,angle=0]{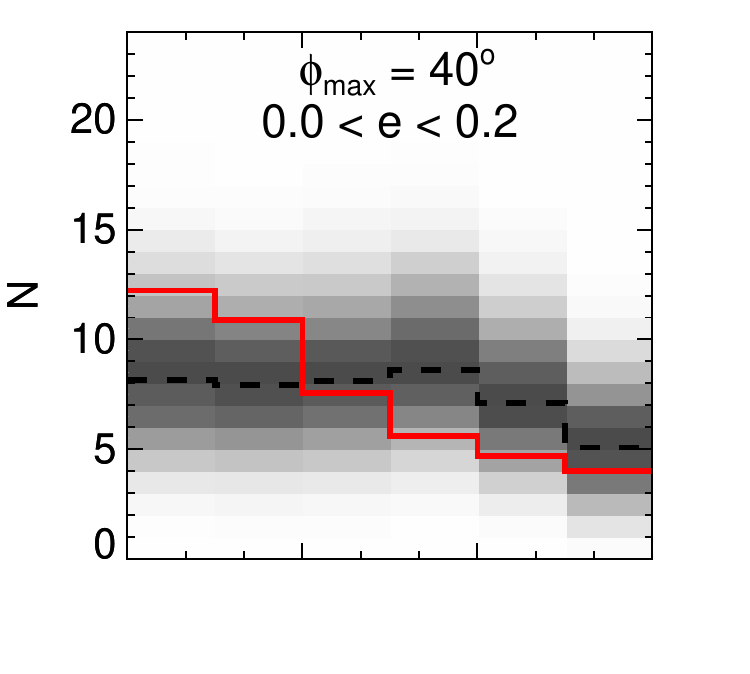}\hskip -0.35in
    \includegraphics[width=1.4in,angle=0]{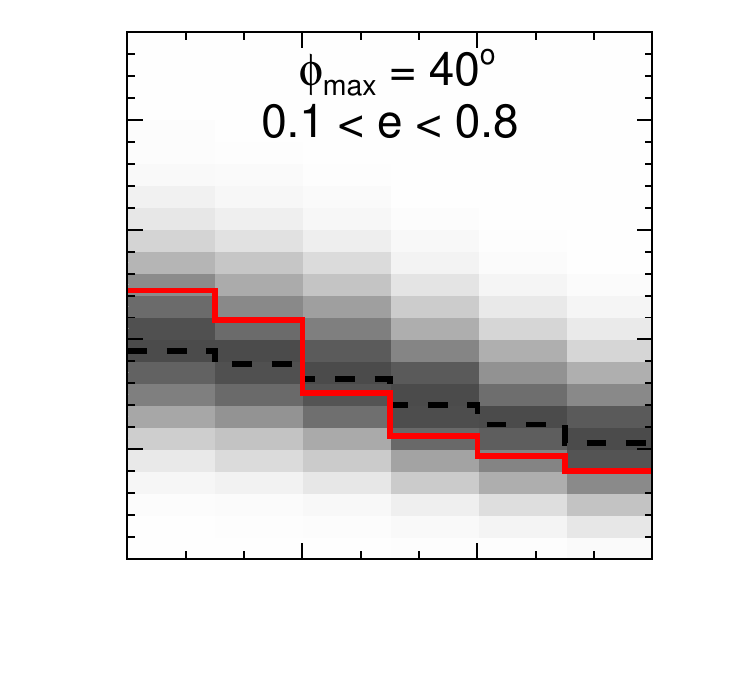}\hskip -0.35in
    \includegraphics[width=1.4in,angle=0]{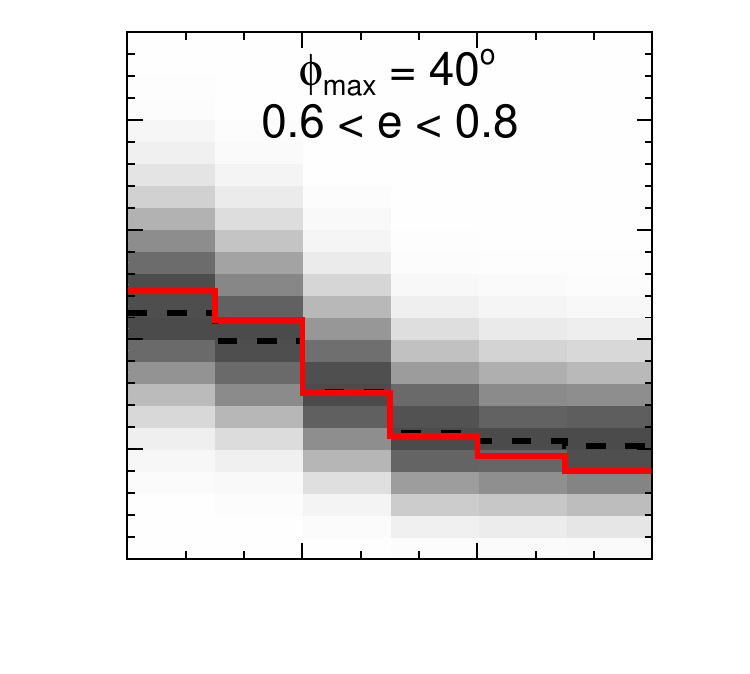}}\vskip -0.24in
  \hskip 0.05in\centerline{
    \includegraphics[width=1.4in,angle=0]{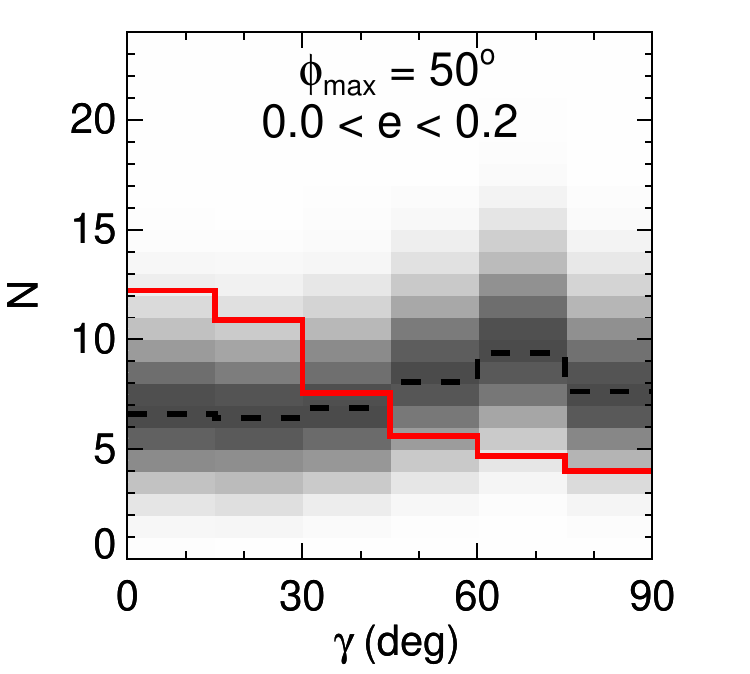}\hskip -0.35in
    \includegraphics[width=1.4in,angle=0]{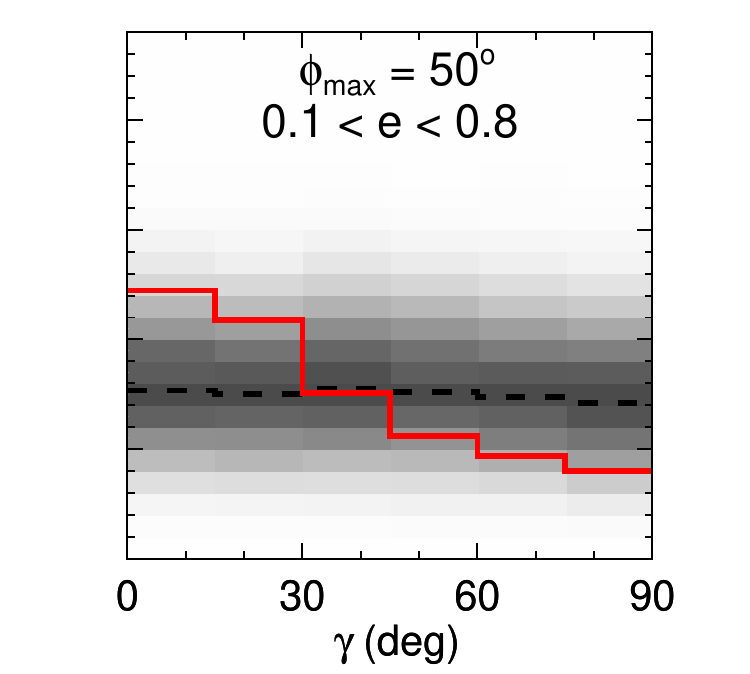}\hskip -0.35in
    \includegraphics[width=1.4in,angle=0]{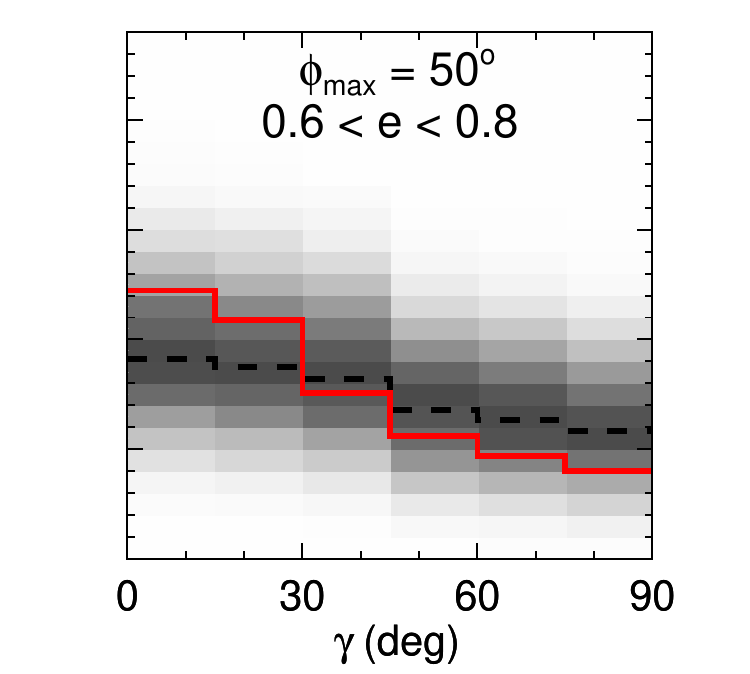}}\vskip -0.05in
  \caption{\normalsize Histograms of simulated $\gamma$ values, same as Figure~\ref{fig:sim-hist-iso}, except here the planet-binary mutual inclinations are uniformly distributed from perfectly aligned ($\phi=0\degree$) up to maximum misalignment of $\phimax$. The most aligned cases are more peaked toward $\gamma=0\degree$ than our observed distribution, and the less aligned cases are too flat, unless eccentricities are also very high. A field binary-like eccentricity distribution (middle column) and $\phimax=30\degree$ provides the best match to the observed $\gamma$ distribution.
  \label{fig:sim-hist-phimax}}
\end{figure}

For the eccentricity distribution, the null hypothesis is that it would resemble the field binary population, which is roughly flat over a wide range of eccentricities \citep{2010ApJS..190....1R}. In addition to this uniform distribution ranging from $0.1<e<0.8$, we also tested very low ($0.0<e<0.2$) and very high ($0.6<e<0.8$), uniformly-distributed eccentricity distributions. These three eccentricity distributions, combined with the ten mutual inclination distributions, resulted in a total of 30 distinct simulations of binary orbital arcs.

\subsubsection{Statistical tests of simulated $\gamma$ distributions \label{sec:stats}}

Figure~\ref{fig:sim-hist-iso} shows distributions of $\gamma$ for isotropic orbital inclinations, as would be expected if the transiting planet orbital plane were independent of the binary orbital plane. Unlike the observed distribution of $\gamma$, these distributions all have minima at low $\gamma$. This is the opposite of the observed distribution that has a minimum at high $\gamma$ and rises to peak at low $\gamma$. Even very high eccentricities are not sufficient to fully reverse the shape of the $\gamma$ distribution under the assumption of isotropic inclinations.

Next, Figure~\ref{fig:sim-hist-phi0} shows distributions of $\gamma$ when the planet and binary orbits are all misaligned by a specific amount. For instance, in the end state of Kozai-Lidov migration binaries and planets would be expected to be misaligned by $\gtrsim$40\degree\ \citep[e.g.,][]{2007ApJ...669.1298F,2011Natur.473..187N}. Distributions of $\gamma$ for mutual inclinations that range from perfectly aligned up to a varying degree of maximum misalignment are shown in Figure~\ref{fig:sim-hist-phimax}. In general, highly-misaligned orbits do not provide a good match to the observed distribution, again regardless of eccentricity. Yet highly-aligned orbits also provide a poor match, peaking more strongly at low $\gamma$ than is observed. 

To quantitatively assess the significance of differences between our observed orbital arcs and simulated data, we performed tests of the cumulative distribution functions of $\gamma$. In order to account for the fact that our orbital arc measurement errors ($\dot{\rho}$ and $\dot{\theta}$) do not necessarily propagate to symmetric errors on $\gamma$, we conducted our tests using $10^4$ Monte Carlo trials. We computed the Kolmogorov-Smirnov (K-S) and Anderson-Darling (A-D) statistics, and the associated K-S probability, for every trial and adopted the mean value obtained for each of the 50 simulated orbit distributions. 

The field binary-like eccentricity distribution with aligned orbits up to $\phimax = 30$\degree\ gave the best K-S statistic ($p=0.81$) and best A-D statistic (0.03). In fact, regardless of the eccentricity distribution tested, this same mutual inclination distribution always gave good K-S ($p>0.55$) and A-D ($<$0.04) statistics. Several other simulated orbit parameters gave acceptable statistics, generally falling into two categories. One is comparable to the best-matching case, except with slightly lower or higher values of $\phimax$, ranging from 20\degree\ to 50\degree. Within this category, the more aligned cases showed a preference for more circular orbits, and the converse was true for less aligned orbits. This follows the intuition that either misalignment or higher eccentricity can skew the $\gamma$ distribution to lower values. The K-S tests ranged from $p=0.10$--0.76 for this category of orbits, and the A-D statistic ranged from 0.06--0.12. 

The other category of simulated orbits that gave good matches to the observed distribution used small and narrowly-distributed mutual inclinations ($\phi_0 = 15$\degree) with a slight preference for eccentric orbits ($p=0.54$ for $0.6<e<0.8$) but acceptable fits for field-like eccentricities ($p=0.30$ for $0.1<e<0.8$). Compared to the other category of simulated orbits, these had larger A-D statistics of 0.14 and 0.25, respectively. This reflects the different sensitivity of the A-D statistic and the fact that this category or simulations deviates from the observed distribution in a different way than the other category. In these cases, the high $\gamma$ tail is matched well, but the simulated distribution does not peak toward $\gamma=0$\degree\ like the simulations that include more well-aligned mutual inclinations.

The worst performing inclination distribution was the assumption of isotropic orbits, where the binary and planet orbits would have no knowledge of each other. The best K-S statistic was for the case of only highly eccentric orbits ($p=0.006$ for $0.6<e<0.8$), and the next best case of field-like eccentricities was much worse ($p=0.0002$). The A-D statistics for the isotropic case were also much worse, ranging from 0.22--0.50.

\begin{figure}
  \hskip 0.05in
  \centerline{
    \includegraphics[width=1.4in]{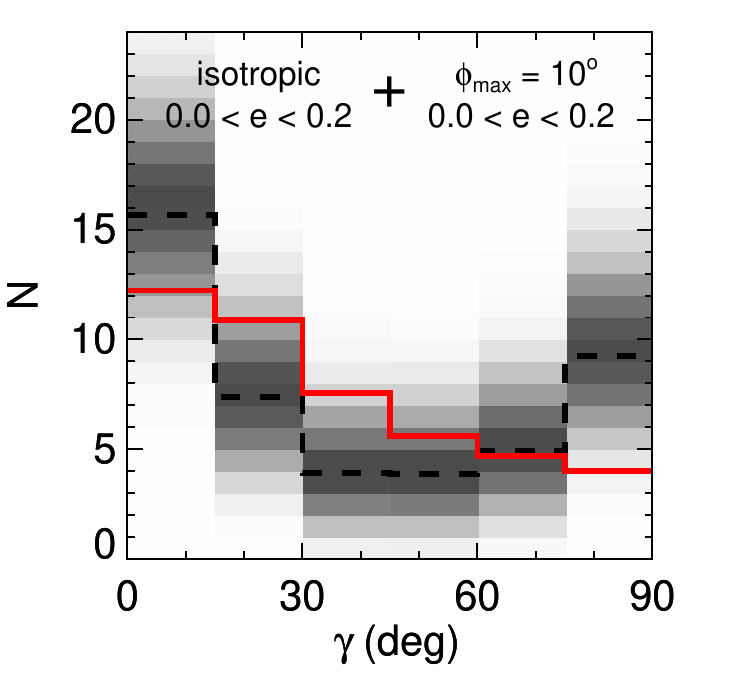}\hskip -0.35in
    \includegraphics[width=1.4in]{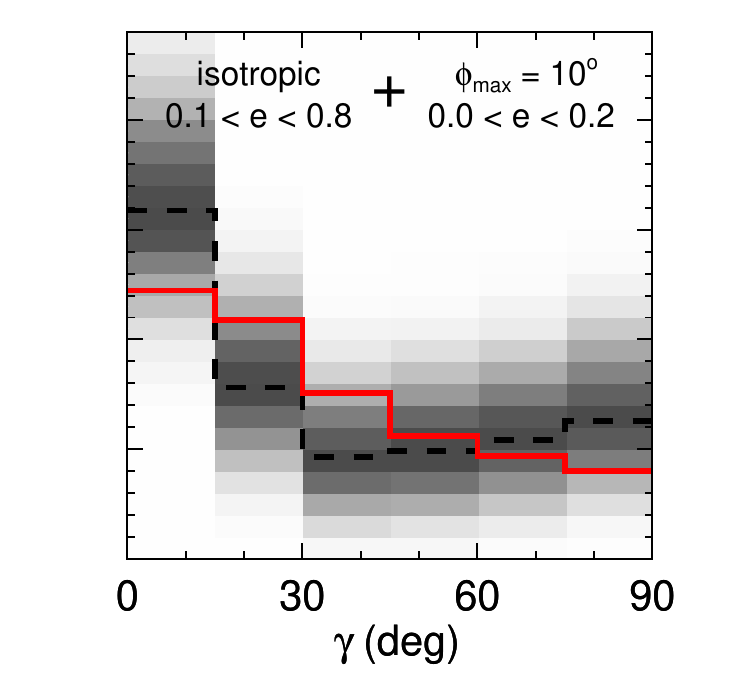}\hskip -0.35in
    \includegraphics[width=1.4in]{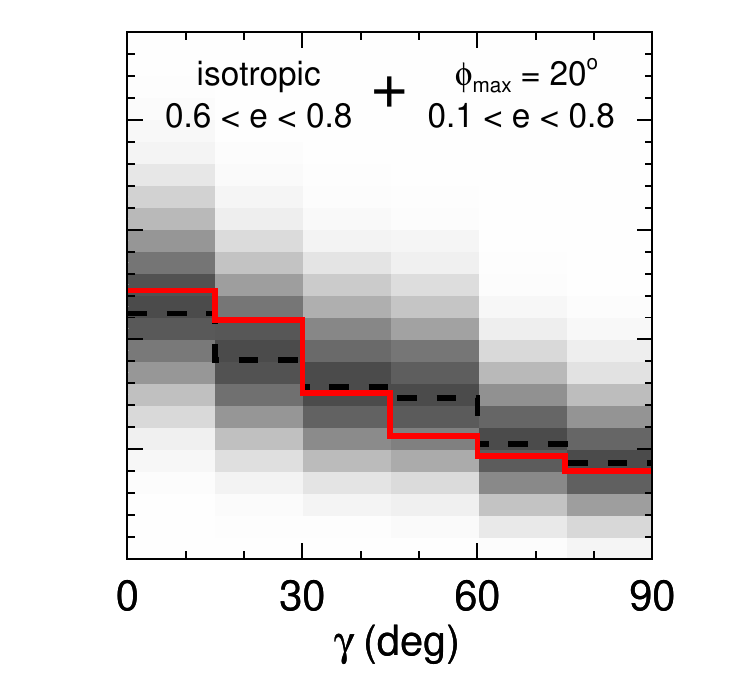}}\vskip -0.05in
  \caption{\normalsize Histograms of simulated $\gamma$ values, same as Figure~\ref{fig:sim-hist-iso}, except here only half of the underlying orbital parameters correspond to isotropic orbits with either low eccentricities (left), field-like eccentricities (middle), or high eccentricities (right). The other half of the population in each case corresponds to the mutual inclination and eccentricity distributions that provide the best match to the observed data. None of the best-matching two-population simulations that include isotropic orbits provides a better match than the best single-population simulations, but their K-S test probabilities ($p=0.27$--0.72) and A-D statistics (0.025--0.035) are acceptable.
  \label{fig:sim-hist-2pop}}
\end{figure}

\subsection{Two-Population Tests}

Although we found many single-population binary orbit simulations that match our observations, none of them included isotropically-distributed binary inclinations. Given that there may well be multiple formation pathways to produce the \Kepler\ planet sample, we have also considered the possibility that two disparate populations could explain our observed distribution of $\gamma$. Two-population models have been proposed to explain other properties of the \Kepler\ sample, as for instance the ``Kepler dichotomy'' has previously been shown to be necessary to explain the number of single- versus multi-transiting systems \citep{2011ApJS..197....8L,2016ApJ...816...66B}. For simplicity, our simulations assume an exact 50/50 split in the population, which is also comparable to the Kepler dichotomy split.

We tested combinations of all three eccentricity distributions and either isotropic plus aligned (up to $\phimax$) or isotropic plus misaligned (by $\phi_0$) orbits. Many of the 81 pairings produced acceptable K-S statistic probabilities, and we shown the best of each isotropic case in Figure~\ref{fig:sim-hist-2pop}. Broadly speaking, the results indicate that if half of the orbits are isotropic and have low or field binary-like eccentricities, then the other half need to have very small mutual inclinations ($\phimax=10$\degree) and eccentricities ($e<0.2$). If the isotropic half has high eccentricity, then the other half still needs to have aligned mutual inclinations ($\phimax=20$\degree) and a field-binary like eccentricity distribution actually provides the best match. Many other pairings produced poor matches to the observations, and the worst ones used the highly-misaligned case ($\phi_0=45$\degree) that corresponds to Kozai-Lidov migration.

Overall, the results of our two-population orbit simulations show that a majority of our sample must show some degree of alignment between planet and binary orbits. If some of the sample has misaligned or isotropic orbits, then the rest need to be even more aligned (and less eccentric) than indicated by the single-population tests. Of all the two-population tests, the best match is actually a combination of highly eccentric orbits with random mutual inclinations and well-aligned orbits with a field binary-like eccentricity distribution.

\section{Discussion}

The most general result of our tests of stellar orbital direction is that the orbital planes of our planet-hosting binary sample are not drawn from a random distribution. More specifically, a significant subset of the sample must have low mutual inclinations in order to explain the observed distribution of orbital arcs. High eccentricities are not required to explain the tendency of stellar companions' orbital motion to be mostly toward or away from the primary star, but high eccentricities are not ruled out as long as there is also still a tendency toward alignment. A single population of mostly-aligned, high-eccentricity orbits provides a perfectly adequate match to the observations, as well as two populations split between high eccentricities with random inclinations and field binary-like eccentricities with well-aligned mutual inclinations. 

In the theoretical interpretation that follows, we make the fundamental assumption that the stars in these systems formed first, and then the planets formed subsequently in circumstellar disks in the presence of the binary companions. Given that these systems only contain two stars, along with their one or more planets, it is implausible that substantial  evolution of the binary orbits has taken place since the formation epoch. At such early times, it is however possible in principle that planets form around one star before the binary companion has evolved onto its current, close-in orbit. This possibility is mostly inconsistent with binary statistics that show that the orbital properties of field binaries are in place even on the pre-main sequence \citep{2018ApJ...854...44M,2019AJ....157..196K}. Therefore, the most likely scenario is that the present-day binary orbits record the architectures of the systems while their planets were forming in their disks. In the following, we consider the various types of orbital configurations that we have inferred from our orbital arc analysis in the context of planet formation theory.

The simplest explanation for the low mutual inclinations that we infer is that close binaries and their disks preferentially begin as (or quickly evolve to) coplanar systems. This explanation is consistent with close-binary formation models based on disk fragmentation \citep{1989ApJ...347..959A,Bonnell:1994ob,2018ApJ...854...44M,2019AJ....158..167T}, in which the gas disks form in the plane of the binary orbit. 

Even if close binaries arise from the migration of much wider original pairs of stars \citep{2012MNRAS.419.3115B,2019MNRAS.484.2341B,2019ApJ...887..232L}, gas-rich disks can warp and align with the binary orbit on timescales short compared to the disk lifetime \citep{2000MNRAS.317..773B}. Whether this is a viable pathway depends on the details of the alignment timescale, which itself depends on the origin of effective viscosity in the disk and the presence of parasitic modes that damp warp excitation.

Another complication in theoretical predictions for low mutual inclinations between binary orbits and circumstellar disks is that even modest initial misalignments between binary, planet, and disk orbital planes can lead to substantial inclination excitation for the planets and disks \citep{2016ApJ...817...30L,Franchini20}. These works consider the evolution of tilt oscillations in the absence of viscous damping. They find that multi-planet systems can have relative inclinations grow, and in some cases planets can evolve onto retrograde orbits. The independent evolution of planets in the system depends on their proximity and relative masses. In some cases the gravitational interactions between tightly packed planets lead to oscillations more like a rigid disk, thus preserving planetary alignment \citep{Petrovich2020}. 

Our sample is well-matched by low mutual inclinations ($\lesssim$30\degree), but purely coplanar orbits (as flat as the solar system, for instance) cannot entirely reproduce our sample. This suggests that multiple processes are at work in shaping the distribution of planet-binary inclinations, some of which may result in fully coplanar systems and some more misaligned. There is good evidence for a fully coplanar configuration in at least the case of KOI-0001~A, better known as the parent star of TrES-2 \citep{2006ApJ...651L..61O}. The low obliquity of TrES-2 is consistent with alignment between the stellar spin and the 0.04-AU planetary orbit\citep{2008ApJ...682.1283W}, and our orbital arc shows no significant PA motion but a 12$\sigma$ detection of edge-on orbital motion. This implies a remarkable alignment of the 240-AU stellar companion's orbit with the hot Jupiter and its host star.

KOI-3444~AB is another notable case highlighting multiple planes of orbital alignment. As discussed in Section~\ref{sec:fp}, it is the only binary in our sample with clear evidence for at least one planet orbiting both of the stars in the system. We detect edge-on orbital motion at 6$\sigma$ but only 2$\sigma$ in the orthogonal direction. This system is thus consistent with alignment between the planetary system around the primary, the planetary system around the secondary, and the binary's orbit that has a projected separation of 98\,AU.

Our results do not rule out high eccentricities for our sample, as long as mutual inclinations are also low in at least a subset of systems. However, the conclusion that planet occurrence rates are highest in the most eccentric binaries is at odds with theoretical expectations. Moreover, it would be inconsistent with both previous investigations of planet occurrence rates \citep{2019arXiv191201699M} and would contradict theoretical predictions for the impact of binaries on planet formation. For one, more eccentric binaries truncate their disks, and thereby their planet-forming mass reservoirs, at smaller radii for a fixed semi-major axis \citep{1994ApJ...421..651A}?. The case of KOI-3158~AB (Kepler-444~AB) illustrates this point well, as it does indeed have a high eccentricity (and is also consistent with being aligned) but its Mars-sized planets are smaller than any other binary system in our sample. Secondly, eccentric stellar companions tend to excite planetesimal velocities, inhibiting core formation \citep{2015ApJ...798...71S}. Lastly, high-eccentricity binaries would tend to drive higher eccentricities in planets through dynamical interactions, leading to more orbital parameter space that is dynamical unstable and thus likely fewer planets \citep{1999AJ....117..621H,2006ApJ...639..423M}.

\section{Summary}

We present the results of a five-year astrometric monitoring campaign studying 45 binary star systems that host \Kepler\ planet candidates, including Kepler-444 that was the first result from our observations \citep{2016ApJ...817...80D}. The overall goal of our observations is to connect planet formation outcomes (like the \Kepler\ sample) to the initial conditions (like disk size and alignment). The fundamental assumption is that stars must form before planets, and thus the planet-forming environment is literally shaped by the stellar orbits that remain unchanged to the present day. One of the most important orbital characteristics, mutual inclination, can only be addressed using a statistical sample as projection effects impede conclusive results on individual systems.

We describe how we selected our sample, which was designed to follow the fastest moving orbits (in angular units) among the binaries reported in the survey of \citet{2016AJ....152....8K}. We reassess the false-positive status of all of our targets and show that they are all consistent with at least one planet orbiting one star in the system. All are consistent with the primary having at least one planet, although in many cases the planet could be orbiting the secondary star instead. One system (KOI-3444~AB) has a transiting planet around each star.

We used Keck/NIRC2 in concert with NGS and LGS AO to obtain multi-epoch imaging and aperture masking observations, with results presented here based on data collected over 2012 to 2017. We derived high-precision relative astrometry from these data and found the orbital arcs to be consistent with linear motion, and gravitationally bound orbits, for our entire sample. Our repeated observations allowed us to validate our astrometric errors and examine them as a function of binary separation. A typical system in our sample has uncertainty in its linear motion of $\sim$0.1\,\masyr, with the best and worst systems ranging from $\approx$0.03\,\masyr\ to $\approx$3\,\masyr.

We performed a number of tests that examine the distribution of orbital velocities measured in our sample, focusing specifically on the direction of the orbital vector. Stellar companions in edge-on orbits will display only motion toward or away from the primary stars, and we find a preponderance of such orbits in our sample that is inconsistent with the null hypothesis of randomly oriented binary orbits at 4.7$\sigma$. Given that our sample was defined by having transiting (edge-on) planetary orbits, this result suggests that planet-binary alignment is common.

In order to test whether high eccentricity could be mimicking orbital alignment, we created simulations that varied these two key unknown orbital parameters. We found that our observed distribution of orbital arcs can only be explained if most of the systems have low mutual inclinations between planet and stellar orbits. Randomly distributed mutual inclinations cannot explain the observed distribution of orbital arcs alone, even with very high eccentricities. A single underlying distribution of orbital parameters in which our binaries have eccentricities that follow field binaries and planet-binary mutual inclinations are distributed uniformly between 0\degree\ and 30\degree\ provides the best match to our observed orbital arcs.

We discuss the implications of widespread planet-binary alignment in the theoretical context of planet formation and circumstellar disk evolution. Our results are consistent with planets in binaries forming preferentially in disks that are moderately well aligned with the binary orbital plane. This finding points to either preferential formation of planets in binaries with primordial co-planar disks, or in systems whose disks are torqued into alignment on timescales faster than the disk dissipation timescale. In either case, modest misalignments for some systems are consistent with the myriad dynamical mechanisms for exciting inclination.

Our observations of planet-hosting binaries are ongoing with a few immediate goals. One is to develop a sample that is not biased toward the fastest moving systems. Our observations began before much of the \Kepler\ planet candidate hosts had high-resolution imaging and before \Gaia\ distances were available. It is now possible to construct a volume-limited sample of \Kepler\ hosts that have by now been thoroughly surveyed for binaries \citep[e.g.,][]{Furlan:2017aa,Ziegler:2017aa}. In the present work we focused on the distance- and mass-independent direction of the orbital vector. With \Gaia\ distances and improved stellar properties, it will be possible to perform orbit tests in physical units (\kms) with accurate constraints on the total system masses. With such a carefully constructed and larger sample, it will be possible to examine trends among subsets of the planetary systems as well. For instance, whether planet size or planet multiplicity is related to the underlying orbital properties, and thereby different planet-forming environments.

With its depth of planetary demographic information, the \Kepler\ sample continues to be one of the most fertile data sets for addressing questions about the formation and evolution of exoplanetary systems. For studies of planet-hosting binary systems, the large distances to most of the \Kepler\ planet hosts presents a serious hindrance to more detailed orbital studies, such as pinpointing the parameters of individual systems. The sample of {\sl TESS} planets orbiting more nearby stars will be complementary to \Kepler\ in this way, as it will open the door to full orbit fits and much closer binary separations.

\section*{Acknowledgements}

We are grateful to the anonymous referee for comments that improved our manuscript.
A.C.R.\ was supported as a 51~Pegasi~b Fellow though the Heising-Simons Foundation.
K.M.K.\ acknowledges support from NASA ATP Grant 80NSSC18K0726.
The data presented herein were obtained at the W.M.\ Keck Observatory, which is operated as a scientific partnership among the California Institute of Technology, the University of California, and the National Aeronautics and Space Administration. The Observatory was made possible by the generous financial support of the W.M.\ Keck Foundation.
This work has made use of data from the European Space Agency (ESA) mission Gaia (\url{https://www.cosmos.esa.int/Gaia}), processed by the Gaia Data Processing and Analysis Consortium (DPAC, \url{https://www.cosmos.esa.int/web/Gaia/dpac/consortium}). Funding for the DPAC has been provided by national institutions, in particular the institutions participating in the Gaia Multilateral Agreement.

\section*{Data Availability}

All of our NIRC2 data are available on the Keck Observatory Archive (KOA), which is operated by the W.\ M.\ Keck Observatory and the NASA Exoplanet Science Institute (NExScI), under contract with the National Aeronautics and Space Administration.



\begin{table*}
\addtocounter{table}{-1}
\caption[]{Relative Astrometry of KOIs from Keck/NIRC2 Adaptive Optics Imaging and Masking.} \label{tbl:keck-full}
\begin{tabular}{llccccl}
\hline
System & \multicolumn{2}{c}{Observation epoch} & Separation & Position angle & $\Delta{m}$ & Filter \\
 & (UT) & (MJD) & (mas) & ($\degree$) & (mag) & \\ \hline
KOI-0001AB & 2012~Jul~6  & 56114.59 & $1105.1\pm0.4    $ & $136.351\pm0.020  $ & $ 2.351\pm0.000  $ & $\Kp     $ \\
KOI-0001AB & 2015~Jun~23 & 57196.55 & $1110.3\pm0.5    $ & $136.369\pm0.022  $ & $ 2.386\pm0.007  $ & $\Kp     $ \\
KOI-0001AB & 2016~Jun~16 & 57555.59 & $1112.7\pm0.5    $ & $136.385\pm0.022  $ & $ 2.347\pm0.008  $ & $\Kp     $ \\\hline
KOI-0042AB & 2012~May~6  & 56053.64 & $1667.2\pm0.7    $ & $ 35.534\pm0.020  $ & $ 1.854\pm0.020  $ & $\Kp     $ \\
KOI-0042AB & 2014~Jul~31 & 56869.36 & $1666.3\pm0.7    $ & $  35.61\pm0.04   $ & $  1.86\pm0.03   $ & $\Kc     $ \\
KOI-0042AB & 2015~Jul~27 & 57230.43 & $1665.2\pm0.8    $ & $ 35.592\pm0.024  $ & $ 1.905\pm0.015  $ & $\Kc     $ \\
KOI-0042AB & 2016~Jun~16 & 57555.57 & $1664.4\pm0.8    $ & $ 35.607\pm0.024  $ & $ 1.838\pm0.020  $ & $\Kc     $ \\\hline
KOI-0112AB & 2014~Jul~31 & 56869.32 & $ 101.1\pm0.6    $ & $ 115.50\pm0.17   $ & $ 1.126\pm0.025  $ & $\Kp     $ \\
KOI-0112AB & 2015~Jul~21 & 57224.51 & $  99.3\pm0.8    $ & $ 114.76\pm0.25   $ & $ 1.125\pm0.022  $ & $\Kp     $ \\
KOI-0112AB & 2016~Jul~15 & 57584.43 & $ 100.7\pm0.9    $ & $  113.5\pm0.5    $ & $ 1.149\pm0.029  $ & $\Kp     $ \\
KOI-0112AB & 2017~Jul~7  & 57941.32 & $ 103.3\pm1.9    $ & $  112.6\pm0.6    $ & $  1.23\pm0.07   $ & $\Kp     $ \\\hline
KOI-0214AB & 2014~Aug~13 & 56882.30 & $  70.9\pm1.6    $ & $  196.2\pm1.3    $ & $  3.71\pm0.10   $ & $\Kp+9  $H \\
KOI-0214AB & 2015~Jul~21 & 57224.54 & $  65.1\pm2.8    $ & $  194.6\pm2.6    $ & $  3.49\pm0.20   $ & $\Kp+9  $H \\\hline
KOI-0227AB & 2013~Jul~28 & 56501.50 & $299.68\pm0.20   $ & $ 68.740\pm0.026  $ & $ 0.021\pm0.001  $ & $\Kp     $ \\
KOI-0227AB & 2014~Jul~30 & 56868.32 & $299.76\pm0.18   $ & $  68.94\pm0.05   $ & $ 0.022\pm0.009  $ & $\Kp     $ \\
KOI-0227AB & 2016~Jul~15 & 57584.42 & $299.73\pm0.18   $ & $  69.33\pm0.03   $ & $ 0.036\pm0.006  $ & $\Kp     $ \\\hline
KOI-0249AB & 2012~Aug~12 & 56151.36 & $4332.3\pm1.8    $ & $ 28.085\pm0.022  $ & $ 0.819\pm0.029  $ & $\Kp     $ \\
KOI-0249AB & 2015~Jul~21 & 57224.33 & $4333.7\pm1.8    $ & $ 28.070\pm0.025  $ & $ 0.790\pm0.027  $ & $\Kp     $ \\\hline
KOI-0270AB & 2014~Jul~31 & 56869.27 & $165.29\pm0.18   $ & $  64.19\pm0.06   $ & $ 0.576\pm0.004  $ & $\Kp     $ \\
KOI-0270AB & 2015~Jun~23 & 57196.54 & $167.47\pm0.16   $ & $  64.31\pm0.04   $ & $ 0.578\pm0.004  $ & $\Kp     $ \\
KOI-0270AB & 2016~Jun~16 & 57555.55 & $170.17\pm0.17   $ & $  64.42\pm0.06   $ & $ 0.543\pm0.009  $ & $\Kp     $ \\
KOI-0270AB & 2016~Jul~15 & 57584.47 & $ 170.3\pm0.4    $ & $  64.46\pm0.08   $ & $ 0.636\pm0.016  $ & $\Kp     $ \\\hline
KOI-0291AB & 2014~Aug~12 & 56881.51 & $  66.2\pm0.3    $ & $ 316.18\pm0.26   $ & $ 1.349\pm0.022  $ & $\Kp+9  $H \\
KOI-0291AB & 2015~Jul~21 & 57224.55 & $  66.0\pm0.3    $ & $  316.8\pm0.3    $ & $ 1.419\pm0.021  $ & $\Kp+9  $H \\
KOI-0291AB & 2017~Jul~7  & 57941.37 & $  64.9\pm0.3    $ & $ 315.58\pm0.27   $ & $ 1.430\pm0.020  $ & $\Kp+9  $H \\\hline
KOI-0588AB & 2012~Aug~12 & 56151.33 & $279.75\pm0.23   $ & $276.561\pm0.020  $ & $ 0.858\pm0.002  $ & $\Kp     $ \\
KOI-0588AB & 2012~Aug~12 & 56151.33 & $279.77\pm0.22   $ & $276.560\pm0.021  $ & $ 0.858\pm0.002  $ & $\Kp     $ \\
KOI-0588AB & 2014~Jul~30 & 56868.45 & $280.71\pm0.25   $ & $ 276.21\pm0.05   $ & $ 0.872\pm0.017  $ & $\Kp     $ \\
KOI-0588AB & 2015~Jul~27 & 57230.51 & $281.05\pm0.14   $ & $ 275.99\pm0.03   $ & $ 0.839\pm0.004  $ & $\Kp     $ \\\hline
KOI-0854AB & 2013~Jul~17 & 56490.53 & $  16.1\pm1.0    $ & $    209\pm5      $ & $  0.30\pm0.23   $ & $\Kp+9  $H \\
KOI-0854AB & 2016~Sep~20 & 57651.29 & $  19.3\pm0.6    $ & $  235.4\pm2.7    $ & $ -0.05\pm0.09   $ & $\Kp+9  $H \\\hline
KOI-0975AB & 2014~Jul~28 & 56866.43 & $ 772.9\pm0.3    $ & $ 129.37\pm0.04   $ & $ 4.244\pm0.010  $ & $\Kc     $ \\
KOI-0975AB & 2015~Jun~23 & 57196.53 & $ 775.7\pm0.5    $ & $ 129.48\pm0.04   $ & $ 4.258\pm0.004  $ & $\Kc     $ \\
KOI-0975AB & 2016~Nov~8  & 57700.26 & $ 780.4\pm0.7    $ & $ 129.57\pm0.03   $ & $ 4.197\pm0.027  $ & $\Kc     $ \\\hline
KOI-1422AB & 2013~Jul~17 & 56490.48 & $ 214.7\pm0.3    $ & $217.016\pm0.020  $ & $ 1.164\pm0.010  $ & $\Kp     $ \\
KOI-1422AB & 2014~Jul~30 & 56868.43 & $214.25\pm0.15   $ & $ 216.94\pm0.04   $ & $ 1.166\pm0.007  $ & $\Kp     $ \\
KOI-1422AB & 2015~Jul~27 & 57230.48 & $213.35\pm0.13   $ & $ 216.82\pm0.06   $ & $ 1.165\pm0.008  $ & $\Kp     $ \\\hline
KOI-1613AB & 2012~Aug~14 & 56153.39 & $211.69\pm0.23   $ & $ 184.49\pm0.04   $ & $ 1.044\pm0.010  $ & $\Kp     $ \\
KOI-1613AB & 2014~Aug~13 & 56882.45 & $209.00\pm0.10   $ & $ 184.60\pm0.05   $ & $ 1.046\pm0.029  $ & $\Kp     $ \\
KOI-1613AB & 2015~Jul~27 & 57230.48 & $207.61\pm0.10   $ & $ 184.49\pm0.04   $ & $ 1.068\pm0.002  $ & $\Kp     $ \\
KOI-1613AB & 2016~Jun~16 & 57555.57 & $206.31\pm0.10   $ & $ 184.53\pm0.03   $ & $ 1.043\pm0.002  $ & $\Kp     $ \\
KOI-1613AB & 2016~Jul~15 & 57584.48 & $206.14\pm0.21   $ & $ 184.60\pm0.04   $ & $ 1.070\pm0.005  $ & $\Kp     $ \\
KOI-1613AB & 2017~Jul~7  & 57941.34 & $204.57\pm0.16   $ & $ 184.43\pm0.05   $ & $ 1.054\pm0.006  $ & $\Kp     $ \\\hline
KOI-1615AB & 2012~Jul~6  & 56114.62 & $  31.8\pm1.6    $ & $  121.5\pm1.6    $ & $  1.81\pm0.10   $ & $\Kp+9  $H \\
KOI-1615AB & 2014~Jul~30 & 56868.55 & $  30.2\pm2.8    $ & $  138.0\pm2.8    $ & $  2.23\pm0.20   $ & $\Kp+9  $H \\
KOI-1615AB & 2014~Nov~30 & 56991.20 & $  23.4\pm2.2    $ & $    139\pm4      $ & $  1.76\pm0.29   $ & $\Kp+9  $H \\
KOI-1615AB & 2016~Sep~20 & 57651.27 & $  17.5\pm0.9    $ & $    145\pm3      $ & $  0.81\pm0.27   $ & $\Kp+9  $H \\\hline
KOI-1619AB & 2012~Jul~6  & 56114.61 & $2066.2\pm0.8    $ & $226.650\pm0.020  $ & $ 2.055\pm0.000  $ & $\Kp     $ \\
KOI-1619AB & 2015~Jul~26 & 57229.56 & $2069.7\pm0.9    $ & $226.594\pm0.022  $ & $ 2.041\pm0.008  $ & $\Kp     $ \\
KOI-1619AB & 2016~Nov~8  & 57700.27 & $2069.9\pm1.6    $ & $226.571\pm0.021  $ & $ 2.111\pm0.014  $ & $\Kp     $ \\\hline
\end{tabular}
\end{table*}

\begin{table*}
\contcaption{} \label{tbl:keck-full-cont}
\begin{tabular}{lcccccl}
\hline
System & \multicolumn{2}{c}{Observation epoch} & Separation & Position angle & $\Delta{m}$ & Filter \\
 & (UT) & (MJD) & (mas) & ($\degree$) & (mag) & \\ \hline
KOI-1681AB & 2013~Jul~17 & 56490.54 & $ 149.9\pm0.6    $ & $141.240\pm0.027  $ & $ 0.093\pm0.006  $ & $\Kp     $ \\
KOI-1681AB & 2014~Jul~29 & 56867.37 & $148.69\pm0.18   $ & $ 141.23\pm0.07   $ & $ 0.082\pm0.008  $ & $\Kp     $ \\
KOI-1681AB & 2015~Jul~27 & 57230.49 & $146.71\pm0.18   $ & $ 141.12\pm0.07   $ & $ 0.095\pm0.007  $ & $\Kp     $ \\\hline
KOI-1725AB & 2014~Jul~31 & 56869.38 & $4056.8\pm1.6    $ & $ 98.582\pm0.029  $ & $ 1.456\pm0.004  $ & $\Kp     $ \\
KOI-1725AB & 2014~Aug~13 & 56882.36 & $4053.6\pm1.6    $ & $ 98.573\pm0.020  $ & $ 1.637\pm0.012  $ & $\Kp     $ \\
KOI-1725AB & 2015~May~28 & 57170.62 & $4050.0\pm2.8    $ & $ 98.548\pm0.023  $ & $ 1.606\pm0.017  $ & $\Kp     $ \\\hline
KOI-1835AB & 2012~Aug~12 & 56151.40 & $  58.0\pm1.9    $ & $  354.3\pm0.9    $ & $ 0.035\pm0.018  $ & $\Kp     $ \\
KOI-1835AB & 2014~Jul~30 & 56868.54 & $  55.6\pm1.1    $ & $  354.2\pm1.2    $ & $  0.00\pm0.03   $ & $\Kp+9  $H \\
KOI-1835AB & 2015~Jul~22 & 57225.43 & $  53.4\pm0.5    $ & $  355.8\pm0.7    $ & $-0.002\pm0.005  $ & $\Kp+9  $H \\
KOI-1835AB & 2016~Nov~7  & 57699.27 & $  52.1\pm0.3    $ & $ 356.17\pm0.21   $ & $ 0.006\pm0.005  $ & $\Kp+9  $H \\
KOI-1835AB & 2017~Jun~28 & 57932.42 & $  51.1\pm1.5    $ & $  356.3\pm1.6    $ & $ 0.027\pm0.013  $ & $\Kp+9  $H \\\hline
KOI-1841AB & 2012~Aug~12 & 56151.45 & $ 307.0\pm0.4    $ & $ 74.196\pm0.022  $ & $ 2.125\pm0.008  $ & $\Kp     $ \\
KOI-1841AB & 2014~Jul~31 & 56869.31 & $309.06\pm0.24   $ & $  74.14\pm0.04   $ & $ 2.109\pm0.007  $ & $\Kp     $ \\
KOI-1841AB & 2015~Jul~27 & 57230.54 & $310.01\pm0.18   $ & $  74.19\pm0.05   $ & $ 2.033\pm0.008  $ & $\Kp     $ \\
KOI-1841AB & 2017~Jul~7  & 57941.35 & $311.63\pm0.14   $ & $  73.92\pm0.05   $ & $ 2.129\pm0.004  $ & $\Kp     $ \\\hline
KOI-1961AB & 2014~Jul~31 & 56869.49 & $ 34.60\pm0.20   $ & $ 257.60\pm0.20   $ & $ 0.155\pm0.008  $ & $\Kp+9  $H \\
KOI-1961AB & 2015~Jul~21 & 57224.39 & $ 36.99\pm0.15   $ & $ 261.02\pm0.29   $ & $ 0.190\pm0.007  $ & $\Kp+9  $H \\
KOI-1961AB & 2016~Nov~7  & 57699.21 & $  40.0\pm0.4    $ & $  263.0\pm0.8    $ & $ 0.234\pm0.028  $ & $\Kp+9  $H \\
KOI-1961AB & 2017~Jul~1  & 57935.50 & $ 42.17\pm0.18   $ & $  267.5\pm0.4    $ & $ 0.183\pm0.011  $ & $\Kp+9  $H \\\hline
KOI-1962AB & 2012~Aug~14 & 56153.38 & $ 117.8\pm0.8    $ & $  113.7\pm0.3    $ & $  0.13\pm0.05   $ & $\Kp     $ \\
KOI-1962AB & 2014~Jul~31 & 56869.34 & $ 121.8\pm0.5    $ & $ 114.03\pm0.07   $ & $ 0.148\pm0.019  $ & $\Kp     $ \\
KOI-1962AB & 2015~Jul~27 & 57230.47 & $ 124.9\pm0.6    $ & $ 114.03\pm0.18   $ & $ 0.132\pm0.009  $ & $\Kp     $ \\
KOI-1962AB & 2016~Jun~16 & 57555.54 & $125.88\pm0.23   $ & $ 114.20\pm0.12   $ & $ 0.116\pm0.007  $ & $\Kp     $ \\
KOI-1962AB & 2016~Jul~15 & 57584.48 & $125.98\pm0.19   $ & $ 114.32\pm0.07   $ & $ 0.130\pm0.004  $ & $\Kp     $ \\\hline
KOI-1964AB & 2012~Jul~6  & 56114.60 & $394.65\pm0.19   $ & $  1.783\pm0.026  $ & $ 1.936\pm0.011  $ & $\Kp     $ \\
KOI-1964AB & 2014~Jul~28 & 56866.44 & $387.68\pm0.23   $ & $  1.178\pm0.021  $ & $  1.83\pm0.05   $ & $\Kp     $ \\
KOI-1964AB & 2015~Jun~23 & 57196.54 & $385.05\pm0.18   $ & $  0.987\pm0.027  $ & $ 1.908\pm0.006  $ & $\Kp     $ \\
KOI-1964AB & 2016~Jun~16 & 57555.56 & $381.80\pm0.17   $ & $  0.691\pm0.028  $ & $ 1.894\pm0.006  $ & $\Kp     $ \\
KOI-1964AB & 2016~Jul~15 & 57584.48 & $381.63\pm0.20   $ & $  0.695\pm0.028  $ & $ 1.937\pm0.010  $ & $\Kp     $ \\\hline
KOI-1977AB & 2012~Aug~12 & 56151.39 & $ 83.61\pm0.12   $ & $   77.5\pm2.8    $ & $ 0.110\pm0.027  $ & $\Kp     $ \\
KOI-1977AB & 2012~Aug~12 & 56151.39 & $  83.7\pm0.3    $ & $   79.6\pm2.8    $ & $ 0.090\pm0.010  $ & $J       $ \\
KOI-1977AB & 2014~Jul~29 & 56867.54 & $ 80.83\pm0.16   $ & $  76.87\pm0.21   $ & $ 0.123\pm0.011  $ & $\Kp+9  $H \\
KOI-1977AB & 2015~Jul~22 & 57225.42 & $  76.0\pm2.4    $ & $   75.5\pm1.2    $ & $ 0.059\pm0.010  $ & $\Kp+9  $H \\
KOI-1977AB & 2016~Jul~15 & 57584.41 & $  75.9\pm2.7    $ & $   75.6\pm1.6    $ & $  0.14\pm0.07   $ & $\Kp+9  $H \\\hline
KOI-2005AB & 2013~Jul~28 & 56501.50 & $  18.1\pm1.0    $ & $    238\pm6      $ & $ -0.40\pm0.20   $ & $\Kp+9  $H \\
KOI-2005AB & 2014~Jul~29 & 56867.50 & $  19.0\pm1.0    $ & $    250\pm4      $ & $  0.24\pm0.16   $ & $\Kp+9  $H \\
KOI-2005AB & 2015~Jul~21 & 57224.38 & $  20.3\pm0.4    $ & $  236.0\pm1.7    $ & $  0.38\pm0.06   $ & $\Kp+9  $H \\
KOI-2005AB & 2016~Nov~7  & 57699.23 & $  22.9\pm0.9    $ & $  226.0\pm2.8    $ & $  0.17\pm0.07   $ & $\Kp+9  $H \\
KOI-2005AB & 2017~Jul~1  & 57935.48 & $  21.0\pm0.7    $ & $    236\pm4      $ & $  0.03\pm0.08   $ & $\Kp+9  $H \\\hline
KOI-2031AB & 2013~Jul~28 & 56501.50 & $  57.3\pm1.2    $ & $  245.7\pm1.1    $ & $  2.36\pm0.07   $ & $\Kp+9  $H \\
KOI-2031AB & 2014~Jul~29 & 56867.51 & $  55.9\pm1.2    $ & $  249.8\pm1.1    $ & $  2.28\pm0.08   $ & $\Kp+9  $H \\
KOI-2031AB & 2015~Jul~22 & 57225.44 & $  54.8\pm1.2    $ & $  244.4\pm1.3    $ & $  2.15\pm0.08   $ & $\Kp+9  $H \\
KOI-2031AB & 2017~Jul~1  & 57935.47 & $  56.1\pm1.3    $ & $  246.6\pm1.3    $ & $  2.40\pm0.09   $ & $\Kp+9  $H \\\hline
KOI-2059AB & 2012~Aug~13 & 56152.41 & $384.30\pm0.17   $ & $289.749\pm0.024  $ & $ 0.156\pm0.001  $ & $\Kp     $ \\
KOI-2059AB & 2014~Jul~31 & 56869.29 & $ 386.6\pm0.4    $ & $ 289.40\pm0.03   $ & $ 0.129\pm0.006  $ & $\Kp     $ \\\hline
KOI-2124AB & 2012~Aug~12 & 56151.41 & $  48.6\pm0.8    $ & $  53.45\pm0.14   $ & $ 0.000\pm0.006  $ & $\Kp+9  $H \\
KOI-2124AB & 2014~Jul~30 & 56868.36 & $  57.0\pm0.6    $ & $   54.3\pm2.2    $ & $ 0.012\pm0.009  $ & $\Kp+9  $H \\
KOI-2124AB & 2017~Jul~1  & 57935.54 & $  63.9\pm0.3    $ & $  53.63\pm0.29   $ & $ 0.000\pm0.004  $ & $\Kp+9  $H \\\hline
KOI-2179AB & 2013~Jul~18 & 56491.52 & $ 134.6\pm0.8    $ & $ 356.13\pm0.17   $ & $ 0.521\pm0.009  $ & $\Kp     $ \\
KOI-2179AB & 2015~Jul~27 & 57230.50 & $132.55\pm0.14   $ & $ 356.05\pm0.08   $ & $ 0.486\pm0.007  $ & $\Kp     $ \\\hline
KOI-2295AB & 2012~Jul~8  & 56116.59 & $2188.1\pm2.3    $ & $ 78.468\pm0.020  $ & $ 0.921\pm0.016  $ & $\Kp     $ \\
KOI-2295AB & 2015~Jul~22 & 57225.35 & $2188.9\pm1.3    $ & $ 78.605\pm0.025  $ & $ 0.697\pm0.028  $ & $\Kp     $ \\
KOI-2295AB & 2016~Jun~16 & 57555.59 & $2186.4\pm1.5    $ & $ 78.612\pm0.029  $ & $  0.68\pm0.03   $ & $\Kp     $ \\\hline
\end{tabular}
\end{table*}

\begin{table*}
\contcaption{} \label{tbl:keck-full-cont2}
\begin{tabular}{lcccccl}
\hline
System & \multicolumn{2}{c}{Observation epoch} & Separation & Position angle & $\Delta{m}$ & Filter \\
 & (UT) & (MJD) & (mas) & ($\degree$) & (mag) & \\ \hline
KOI-2418AB & 2014~Jul~30 & 56868.37 & $ 104.9\pm1.1    $ & $    2.6\pm0.6    $ & $  2.58\pm0.07   $ & $\Kp+9  $H \\
KOI-2418AB & 2015~Jul~21 & 57224.57 & $ 105.1\pm1.0    $ & $    1.1\pm0.6    $ & $  2.69\pm0.07   $ & $\Kp+9  $H \\
KOI-2418AB & 2017~Jul~7  & 57941.37 & $ 100.8\pm1.1    $ & $    2.5\pm0.6    $ & $  2.76\pm0.06   $ & $\Kp+9  $H \\\hline
KOI-2542AB & 2013~Jul~18 & 56491.45 & $ 764.3\pm0.4    $ & $ 28.703\pm0.021  $ & $ 0.973\pm0.020  $ & $\Kp     $ \\
KOI-2542AB & 2014~Jul~30 & 56868.46 & $ 764.3\pm0.4    $ & $ 28.682\pm0.026  $ & $ 0.946\pm0.022  $ & $\Kp     $ \\
KOI-2542AB & 2015~Jul~27 & 57230.52 & $ 763.7\pm0.4    $ & $ 28.638\pm0.023  $ & $ 0.937\pm0.013  $ & $\Kp     $ \\\hline
KOI-2672AB & 2013~Aug~7  & 56511.40 & $ 642.3\pm0.3    $ & $306.437\pm0.027  $ & $ 3.890\pm0.006  $ & $\Kp     $ \\
KOI-2672AB & 2014~Jul~28 & 56866.56 & $ 640.2\pm0.5    $ & $ 306.44\pm0.07   $ & $ 3.915\pm0.017  $ & $\Kp     $ \\
KOI-2672AB & 2015~Jul~27 & 57230.44 & $ 637.9\pm0.4    $ & $ 306.58\pm0.03   $ & $ 3.894\pm0.018  $ & $\Kp     $ \\\hline
KOI-2705AB & 2014~Jul~30 & 56868.34 & $1889.5\pm0.8    $ & $303.925\pm0.022  $ & $ 2.655\pm0.014  $ & $\Kp     $ \\
KOI-2705AB & 2015~Jul~21 & 57224.52 & $1888.0\pm1.1    $ & $303.904\pm0.023  $ & $ 2.532\pm0.016  $ & $\Kp     $ \\
KOI-2705AB & 2016~Nov~9  & 57701.26 & $1887.6\pm0.9    $ & $ 303.87\pm0.04   $ & $  2.69\pm0.03   $ & $\Kp     $ \\\hline
KOI-2733AB & 2014~Jul~29 & 56867.52 & $ 108.2\pm0.4    $ & $ 293.98\pm0.10   $ & $  0.06\pm0.03   $ & $\Kp     $ \\
KOI-2733AB & 2015~Jul~21 & 57224.50 & $106.02\pm0.27   $ & $ 294.59\pm0.14   $ & $-0.007\pm0.020  $ & $\Kp     $ \\
KOI-2733AB & 2016~Nov~15 & 57707.21 & $ 102.5\pm0.3    $ & $  295.1\pm0.5    $ & $ 0.030\pm0.020  $ & $\Kp     $ \\\hline
KOI-2790AB & 2014~Jul~31 & 56869.33 & $254.34\pm0.26   $ & $ 134.88\pm0.03   $ & $ 0.515\pm0.007  $ & $\Kp     $ \\
KOI-2790AB & 2015~Jul~21 & 57224.48 & $253.46\pm0.14   $ & $ 135.01\pm0.03   $ & $ 0.514\pm0.002  $ & $\Kp     $ \\
KOI-2790AB & 2016~Jul~15 & 57584.44 & $253.04\pm0.18   $ & $ 135.03\pm0.05   $ & $ 0.538\pm0.010  $ & $\Kp     $ \\
KOI-2790AB & 2017~Jul~7  & 57941.32 & $252.32\pm0.17   $ & $ 135.12\pm0.07   $ & $ 0.538\pm0.017  $ & $\Kp     $ \\\hline
KOI-3158AB & 2015~Jun~22 & 57195.53 & $1842.3\pm0.7    $ & $252.839\pm0.020  $ & $ 2.041\pm0.009  $ & $\Kc     $ \\
KOI-3158AB & 2015~Jul~21 & 57224.46 & $1841.7\pm0.8    $ & $252.827\pm0.025  $ & $ 2.139\pm0.008  $ & $\Kc     $ \\
KOI-3158AB & 2016~Jun~16 & 57555.64 & $1841.5\pm0.8    $ & $252.832\pm0.022  $ & $ 2.163\pm0.028  $ & $\Kc     $ \\\hline
KOI-3255AB & 2014~Jul~29 & 56867.55 & $181.77\pm0.20   $ & $ 336.04\pm0.04   $ & $ 0.115\pm0.004  $ & $\Kp     $ \\
KOI-3255AB & 2015~Jul~21 & 57224.48 & $181.23\pm0.09   $ & $ 336.05\pm0.04   $ & $ 0.115\pm0.006  $ & $\Kp     $ \\
KOI-3255AB & 2016~Nov~15 & 57707.23 & $180.74\pm0.13   $ & $ 335.91\pm0.06   $ & $ 0.108\pm0.003  $ & $\Kp     $ \\\hline
KOI-3284AB & 2014~Jul~29 & 56867.49 & $438.95\pm0.19   $ & $ 192.82\pm0.03   $ & $ 2.043\pm0.013  $ & $\Kp     $ \\
KOI-3284AB & 2015~Jul~21 & 57224.46 & $438.43\pm0.28   $ & $192.787\pm0.022  $ & $ 2.036\pm0.011  $ & $\Kp     $ \\\hline
KOI-3444AB & 2014~Aug~13 & 56882.29 & $1083.2\pm0.4    $ & $  9.736\pm0.026  $ & $ 2.466\pm0.010  $ & $\Kp     $ \\
KOI-3444AB & 2015~Jul~26 & 57229.50 & $1085.1\pm0.5    $ & $  9.732\pm0.023  $ & $ 2.476\pm0.019  $ & $\Kp     $ \\
KOI-3444AB & 2016~Jun~16 & 57555.61 & $1087.7\pm0.5    $ & $  9.684\pm0.022  $ & $ 2.418\pm0.009  $ & $\Kp     $ \\\hline
KOI-3892AB & 2014~Aug~12 & 56881.47 & $ 117.2\pm1.5    $ & $  339.6\pm0.9    $ & $  4.34\pm0.11   $ & $\Kp+9  $H \\
KOI-3892AB & 2015~Jul~22 & 57225.37 & $ 124.5\pm2.8    $ & $  344.1\pm1.2    $ & $  4.14\pm0.16   $ & $\Kp+9  $H \\\hline
KOI-3991AB & 2014~Jul~31 & 56869.55 & $ 202.8\pm0.6    $ & $ 111.51\pm0.16   $ & $ 1.547\pm0.017  $ & $\Kp     $ \\
KOI-3991AB & 2015~Jul~21 & 57224.44 & $ 201.5\pm0.4    $ & $ 111.57\pm0.03   $ & $ 1.501\pm0.010  $ & $\Kp     $ \\
KOI-3991AB & 2016~Nov~15 & 57707.22 & $199.20\pm0.12   $ & $ 111.85\pm0.09   $ & $ 1.503\pm0.004  $ & $\Kp     $ \\\hline
KOI-4032AB & 2013~Aug~7  & 56511.43 & $ 125.3\pm0.8    $ & $   30.7\pm0.5    $ & $  3.03\pm0.05   $ & $\Kp+9  $H \\
KOI-4032AB & 2014~Jul~31 & 56869.43 & $ 127.1\pm1.1    $ & $   30.6\pm0.5    $ & $  3.00\pm0.07   $ & $\Kp+9  $H \\\hline
KOI-4097AB & 2013~Aug~7  & 56511.47 & $ 178.0\pm2.9    $ & $   14.8\pm0.4    $ & $  3.62\pm0.04   $ & $\Kp     $ \\
KOI-4097AB & 2014~Jul~28 & 56866.51 & $ 177.6\pm1.0    $ & $   16.2\pm0.3    $ & $ 3.749\pm0.027  $ & $\Kp     $ \\
KOI-4097AB & 2015~May~28 & 57170.59 & $ 176.1\pm2.8    $ & $   17.6\pm0.4    $ & $  3.62\pm0.04   $ & $\Kp     $ \\
KOI-4097AB & 2016~Jun~16 & 57555.62 & $ 177.0\pm0.8    $ & $   18.5\pm0.5    $ & $  3.53\pm0.06   $ & $\Kp     $ \\
KOI-4097AB & 2017~Jul~7  & 57941.34 & $ 173.4\pm0.5    $ & $  21.37\pm0.13   $ & $  3.58\pm0.05   $ & $\Kp     $ \\\hline
KOI-4184AB & 2013~Aug~7  & 56511.46 & $205.87\pm0.09   $ & $223.377\pm0.021  $ & $ 0.054\pm0.001  $ & $\Kp     $ \\
KOI-4184AB & 2014~Jul~31 & 56869.29 & $206.21\pm0.09   $ & $ 223.27\pm0.03   $ & $ 0.027\pm0.006  $ & $\Kp     $ \\
KOI-4184AB & 2016~Jul~15 & 57584.43 & $206.76\pm0.10   $ & $ 223.31\pm0.04   $ & $ 0.028\pm0.005  $ & $\Kp     $ \\
KOI-4184AB & 2017~Jul~7  & 57941.33 & $207.03\pm0.09   $ & $223.045\pm0.028  $ & $ 0.040\pm0.003  $ & $\Kp     $ \\\hline
KOI-4252AB & 2014~Aug~11 & 56880.45 & $ 43.03\pm0.08   $ & $ 348.30\pm0.10   $ & $ 0.467\pm0.007  $ & $\Kp+9  $H \\
KOI-4252AB & 2016~Sep~20 & 57651.25 & $ 49.47\pm0.20   $ & $ 339.35\pm0.19   $ & $ 0.422\pm0.014  $ & $\Kp+9  $H \\
KOI-4252AB & 2017~Jun~29 & 57933.38 & $  51.8\pm0.7    $ & $  337.0\pm0.8    $ & $  0.47\pm0.04   $ & $\Kp+9  $H \\\hline
\end{tabular}
\end{table*}

\label{lastpage}
\end{document}